\pdfoutput=1

\documentclass[10pt]{article}

\usepackage{authblk}                
\usepackage[square,numbers]{natbib} 

\usepackage{amsmath,amssymb}                

\usepackage{listings}    
\usepackage{graphicx}    
\usepackage{subcaption}  
\usepackage{algorithm}   
\usepackage{algpseudocode}

\usepackage{booktabs}    
\usepackage{hyperref}    
\usepackage{xurl}        
\usepackage[htt]{hyphenat} 
\usepackage{microtype} 
\DisableLigatures{}    

\usepackage{caption}
\usepackage{setspace}
\usepackage{multirow}
\usepackage{enumerate}
\usepackage{pdflscape}
\usepackage{pgfplots} 

\usepackage{xcolor}
\definecolor{codegreen}{rgb}{0.25,0.5,0.35}
\definecolor{codegray}{rgb}{0.5,0.5,0.5}
\definecolor{codepurple}{rgb}{0.6,0,0}
\definecolor{backcolour}{rgb}{0.95,0.95,0.92}
\definecolor{colorstring}{rgb}{0.5,0,0.35}
\definecolor{rltred}{rgb}{0.5,0,0}
\definecolor{rltgreen}{rgb}{0,0.5,0}
\definecolor{rltblue}{rgb}{0,0,0.5}
\definecolor{DarkGreen}{rgb}{0.00,0.60,0.00}
\definecolor{ScarletRed}{rgb}{0.80,0.00,0.00}
\definecolor{blizzardblue}{rgb}{0.67, 0.9, 0.93}
\definecolor{green-yellow}{rgb}{0.68, 1.0, 0.18}
\definecolor{dkgreen}{rgb}{0,0.6,0}
\definecolor{gray}{rgb}{0.5,0.5,0.5}
\definecolor{mauve}{rgb}{0.58,0,0.82}
\definecolor{lightgrey}{rgb}{0.90,0.90,0.90}
\definecolor{grey}{gray}{0.75}
\definecolor{light-gray}{gray}{0.80}
\definecolor{codeblue}{rgb}{0.13,0.13,1}

\usepackage[table]{xcolor}

\lstdefinelanguage{Kotlin}{
    keywords={package, import, class, object, trait, fun, val, var, if, else, return, null, try, catch, finally, data, open, for},
    keywordstyle=\color{codeblue}\ttfamily,
    ndkeywords={@GetMapping, @RequestParam, @PathVariable, @RestController, ResponseEntity, Long, Int, String, LocalDateTime, DateTimeFormatter, TimeAgoData, List},
    ndkeywordstyle=\color{codepurple}\ttfamily,
    sensitive=true,
    comment=[l]{//},
    morecomment=[s]{/*}{*/},
    commentstyle=\color{codegreen}\ttfamily,
    stringstyle=\color{red}\ttfamily,
    morestring=[b]",
    morestring=[b]'
}

\lstdefinestyle{mystyle}{
    escapechar=©, 
    language=Python,
    backgroundcolor=\color{backcolour},
    basicstyle=\footnotesize\ttfamily,
    identifierstyle=\footnotesize\ttfamily,
    commentstyle=\color{codegreen},
    keywordstyle=\color{colorstring}\bfseries,
    morekeywords={OR, AND},
    numberstyle=\ttfamily\color{codegray},
    stringstyle=\ttfamily\color{DarkGreen},
    breakatwhitespace=false,
    breaklines=true,
    captionpos=b,
    keepspaces=true,
    numbers=left, 
    numbersep=2pt,
    showspaces=false,
    showstringspaces=false,
    showtabs=false,
    tabsize=2
}
\usepackage{braket}		
\usepackage{mathtools}	

\usepackage{pifont}

\usepackage{xspace}
\newcommand{\evo}{{\sc EvoMaster}\xspace}

\newcommand{\method}{{\sc FlakyCatch}\xspace}

\newcommand{\aratrl}{ARAT-RL\xspace}
\newcommand{\emrest}{EmRest\xspace}
\newcommand{\llamaresttest}{LLamaRestTest\xspace}
\newcommand{\restler}{RESTler\xspace}
\newcommand{\schemathesis}{Schemathesis\xspace}
\newcommand{\deeprest}{DeepRest\xspace}
\newcommand{\apirl}{APIRL\xspace}
\newcommand{\apif}{APIF\xspace}
\newcommand{\astra}{ASTRA\xspace}
\newcommand{\autoresttest}{AutoRestTest\xspace}
\newcommand{\bboxrt}{bBOXRT\xspace}
\newcommand{\kat}{KAT\xspace}
\newcommand{\logiaagent}{LogiaAgent\xspace}
\newcommand{\miner}{MINER\xspace}
\newcommand{\morest}{Morest\xspace}
\newcommand{\nautilus}{Nautilus\xspace}
\newcommand{\openapifuzzer}{OpenAPI-Fuzzer\xspace}
\newcommand{\raft}{RAFT\xspace}
\newcommand{\restct}{RestCT\xspace}
\newcommand{\restest}{RESTest\xspace}
\newcommand{\resttestgen}{RestTestGen\xspace}
\newcommand{\voapi}{VoAPI2\xspace}
\newcommand{\wuppiefuzz}{WuppieFuzz\xspace}

\newcommand{\fuzzenv}{\textit{FuzzEnv}\xspace}
\newcommand{\execenv}{\textit{ExecEnv}\xspace}

\newcommand{\bibliothek}{\emph{bibliothek}\xspace}
\newcommand{\blogapi}{\emph{blogapi}\xspace}
\newcommand{\catwatch}{\emph{catwatch}\xspace}
\newcommand{\cwaverification}{\emph{cwa-verification}\xspace}
\newcommand{\ercrestservice}{\emph{erc20-rest-service}\xspace}
\newcommand{\familiebasak}{\emph{familie-ba-sak}\xspace}
\newcommand{\featuresservice}{\emph{features-service}\xspace}
\newcommand{\genomenexus}{\emph{genome-nexus}\xspace}
\newcommand{\gestaohospital}{\emph{gestaohospital}\xspace}
\newcommand{\httppatchspring}{\emph{http-patch-spring}\xspace}
\newcommand{\languagetool}{\emph{languagetool}\xspace}
\newcommand{\market}{\emph{market}\xspace}
\newcommand{\microcks}{\emph{microcks}\xspace}
\newcommand{\ocvn}{\emph{ocvn}\xspace}
\newcommand{\ohsomeapi}{\emph{ohsome-api}\xspace}
\newcommand{\paypublicapi}{\emph{pay-publicapi}\xspace}
\newcommand{\personcontroller}{\emph{person-controller}\xspace}
\newcommand{\proxyprint}{\emph{proxyprint}\xspace}
\newcommand{\quartzmanager}{\emph{quartz-manager}\xspace}
\newcommand{\reservationsapi}{\emph{reservations-api}\xspace}
\newcommand{\restncs}{\emph{rest-ncs}\xspace}
\newcommand{\restnews}{\emph{rest-news}\xspace}
\newcommand{\restscs}{\emph{rest-scs}\xspace}
\newcommand{\restcountries}{\emph{restcountries}\xspace}
\newcommand{\scoutapi}{\emph{scout-api}\xspace}
\newcommand{\sessionservice}{\emph{session-service}\xspace}
\newcommand{\springactuatordemo}{\emph{spring-actuator-demo}\xspace}
\newcommand{\springbatchrest}{\emph{spring-batch-rest}\xspace}
\newcommand{\springecommerce}{\emph{spring-ecommerce}\xspace}
\newcommand{\springrestexample}{\emph{spring-rest-example}\xspace}
\newcommand{\swaggerpetstore}{\emph{swagger-petstore}\xspace}
\newcommand{\tiltaksgjennomforing}{\emph{tiltaksgjennomforing}\xspace}
\newcommand{\trackingsystem}{\emph{tracking-system}\xspace}
\newcommand{\usermanagement}{\emph{user-management}\xspace}
\newcommand{\webgoat}{\emph{webgoat}\xspace}
\newcommand{\youtubemock}{\emph{youtube-mock}\xspace}

\newcommand{\Time}{\emph{Time}\xspace}
\newcommand{\Rand}{\emph{Rand}\xspace}
\newcommand{\Crypt}{\emph{Crypt}\xspace}
\newcommand{\Unord}{\emph{Unord}\xspace}
\newcommand{\RunMsg}{\emph{RunMsg}\xspace}
\newcommand{\DynState}{\emph{State}\xspace}
\newcommand{\Env}{\emph{Env}\xspace}
\newcommand{\Unk}{\emph{Unk}\xspace}
\newcommand{\GenErr}{\emph{GenErr}\xspace}

\newcommand{\rqA}{How frequently do flaky white-box and black-box tests exist in REST APIs?}
\newcommand{\rqB}{What are the primary sources of flakiness in REST APIs?}
\newcommand{\rqC}{How effective is \method in detecting and mitigating flakiness in tests?}
\usepackage{tcolorbox}
\newtcolorbox{resultsbox}[1][]{
    colframe=gray!100,
    colback=white!100,
    coltitle=white,
    title=#1,
    boxsep=1pt,
    left=2pt,
    right=2pt,
    top=2pt,
    bottom=2pt
}

\newenvironment{results}[1][]{
    \begin{resultsbox}[#1]
}{
    \end{resultsbox}
}
\usepackage{ifthen}
\newboolean{showcomments}
\setboolean{showcomments}{true} 

\ifthenelse{\boolean{showcomments}}{
    \newcommand{\nbc}[3]{
            {\colorbox{#3}{\bfseries\sffamily\scriptsize\textcolor{white}{#1}}}
            {\textcolor{#3}{\sf\small$\langle$\textit{#2}$\rangle$}}}
    
}{
    \newcommand{\nbc}[3]{}

}

\newcommand{\squeezeup}{\vspace{-2.5mm}}
\newcommand{\squeezeupfigureAbove}{\vspace{-0.6cm}}
\newcommand{\squeezeupfigureAfter}{\vspace{-0.5cm}}

\usepackage{geometry}
\title{
Detecting and Mitigating Flakiness in REST API Fuzzing
}

\author[1]{Man Zhang}
\author[1]{Chongyang Shen}
\author[2]{Andrea Arcuri}
\author[1]{Tao Yue}

\affil[1]{Beihang University}
\affil[2]{Kristiania University of Applied Sciences and Oslo Metropolitan University}

\date{}

\begin{document}
	
\maketitle

\begin{abstract}
	Test flakiness is a common problem in industry, which hinders the reliability of automated build and testing workflows.
Most existing research on test flakiness has primarily focused on unit and small-scale integration tests. In contrast, flakiness in system-level testing such as REST APIs are comparatively under-explored.
A large body of literature has been dedicated to the topic of fuzzing REST APIs, whereas relatively little attention has been paid to detecting and possibly mitigating negative effects of flakiness in this context.
To fill this major gap, in this paper, we study the flakiness of tests generated by one of the popularly applied REST API fuzzer in the literature, namely \evo, conduct empirical studies with a corpus of 36 REST APIs to understand flakiness of REST APIs. Based on the results of the empirical studies, we categorize and analyze flakiness sources by inspecting near 3000 failing tests. Based on the understanding, we propose \method to detect and mitigate flakiness in REST APIs and empirically evaluate its performance.
Results show that \method is effective in detecting and handling flakiness in tests generated by white-box and black-box fuzzers.
\end{abstract}

{\bf Keywords}: 
REST API, Flaky Test, API Fuzzing

\section{Introduction}
Flaky tests are automated tests that exhibit non-deterministic behavior, i.e., producing different outcomes when executed under the same conditions~\cite{luo2014empirical}. A rich body of literature has investigated flaky tests and identified their negative impact on developer productivity, trust in test suites, and continuous integration pipelines~\cite{parry2021survey, leesatapornwongsa2022flakerepro, gruber2024automatic, li2025hiflaky, moran2019debugging, osikowicz2025empirically}. Based on a recent survey conducted within the BMW Group~\cite{gruber2022survey}, flaky tests are both common and severe in industrial settings, which reinforces the need for systematic approaches to detect and mitigate flakiness.

While prior work primarily focus on conventional test suites, flakiness in REST API testing remains relatively under-explored. REST API tests can be specified as sequences of requests interacting with stateful services. As a result, flakiness occurs when identical request sequences yield different responses across repeated executions, despite no deliberate changes to the system under test (SUT). Compared to traditional test settings, REST APIs are typically more susceptible to such behavior due to their stateful interactions, distributed nature, and dependence on runtime environments.

This challenge is particularly critical in the context of API fuzzing, a widely adopted technique for automatically generating and executing API requests to uncover faults. Existing approaches predominantly operate in black-box settings, leveraging API specifications (e.g., OpenAPI~\cite{openapispec}) or request–response analysis, while some exploit white-box execution information to guide test generation. A variety of REST API testing techniques have been proposed, such as \apif~\cite{wang2024beyond}, \autoresttest~\cite{kim2025autoresttest}, \deeprest~\cite{corradini2024deeprest}, \emrest~\cite{xu2025effective}, \llamaresttest~\cite{kim2025llamaresttest}, \logiaagent~\cite{zhang2025logiagent}, \nautilus~\cite{deng2023nautilus}, \restest~\cite{martinLopez2021Restest}, \resttestgen~\cite{viglianisi2020resttestgen}, and \wuppiefuzz~\cite{rooijakkers2025wuppiefuzz}. However, the impact of flakiness on API fuzzing remains largely unexplored, particularly in terms of how frequently it occurs and what causes it.

To address this gap, in this paper, we aim to systematically investigate flakiness in REST API fuzzing by first addressing the following two research questions (RQs):
{\bf RQ1}: \textit{\rqA}
{\bf RQ2}: \textit{\rqB}
To answer these RQs, we conducted an empirical study on 36 real-world REST APIs using the fuzzer \evo~\cite{arcuri2025tool} in both its \emph{black-box} and \emph{white-box} modes, under two different runtime environments.
We chose \evo~\cite{arcuri2017restful} because it is the state-of-the-art tool that support both black-box and white-box testing. 
Furthermore, many academic REST API fuzzing prototypes only make HTTP calls and are unable to generate executable test suites with assertions (e.g., in JUnit or Pytest format), so they would had been unusable for this study. 

Our analysis reveals that flakiness is widespread: flaky tests were observed in 31 out of 36 APIs across both environments. White-box fuzzing exhibits flaky behavior more frequently than black-box fuzzing, suggesting that it exposes non-deterministic behaviors that remain hidden in black-box settings. At the same time, white-box fuzzing tends to more consistently expose failures, whereas black-box fuzzing more often leads to intermittently failing tests.
To better understand the root causes of flakiness, we further conducted a manual analysis of nearly 3,000 failing tests and identified nine categories of flakiness sources. Among these, the most prominent ones are runtime environment variations, runtime-dependent messages, and stateful resources, highlighting the strong influence of external conditions and system state on REST API behavior.

Based on these insights, we propose \method, a novel approach for detecting and handling flakiness in REST API fuzzing. \method employs a re-execution-based detection mechanism augmented with lightweight inference to efficiently identify flaky tests. Detected flaky tests are then automatically post-processed to improve the stability and reusability of generated test suites. In particular, \method preserves the structural and behavioral coverage of flaky tests while mitigating their non-deterministic failures. 
Finally, we evaluate \method by addressing
{\bf RQ3}: \textit{\rqC}
Our results show that \method effectively identifies flaky tests and improves the stability of test suites.

\textit{Contributions.} 
This paper first presents an empirical study of flakiness in REST API testing, based on 36 real-world APIs and fuzzing-generated test cases under both black-box and white-box settings. We also introduce a manually labeled dataset of failing test cases, providing a valuable resource for benchmarking and further research. From this analysis, we derive a taxonomy of nine distinct sources of flakiness and propose \method, a novel technique for detecting and mitigating selected flakiness issues in REST API fuzzing. All experimental artifacts are made publicly available to support reproducibility and serve as a baseline for future work.

\textit{Structure.} 
In Section~\ref{sec:background}, we briefly discuss flaky tests and \evo, followed by the related work in Section~\ref{sec:relatedwork}. In Section~\ref{sec:empiricalStudy}, we present the emprical study for answering RQ1 and RQ2. In Section~\ref{sec:flakyhandling}, we present \method and its evaluation. Section \ref{sec:threats} provides threats to validity and Section~\ref{sec:conclusions} concludes the paper.

\squeezeup
\section{Background}
\label{sec:background}
In this section, we briefly introduce the concept of flakiness in software testing of REST APIs.
Also, we provide more information on the used fuzzer in this study, i.e., \evo.
\subsection{Flaky Test}
A flaky test may pass in one execution and fail in another, even when it exercises the same code under the same environment and test input.
Primary causes of flaky tests are related to various factors, 
such as asynchronous waits, concurrency issues, test order dependencies, dependencies on external resources, time, randomness, and algorithmic non-determinism~\cite{luo2014empirical,bell2018deflaker,eck2019understanding,parry2021survey}.

In the context of REST API testing, a test can be regarded as a sequence of requests. 
Therefore, a flaky test is reflected as the same sequence of requests producing different responses when executed on the same code. 
The example shown in Figure~\ref{fig:flaky-example} illustrates a \texttt{GET} endpoint that estimates a price based on a given \texttt{base} value provided as a request parameter.
The endpoint introduces a random jitter by generating a random integer using \texttt{randomInt(10)}, which is then added to the base value to compute the total price.
As a result, even when the same request is repeatedly sent with identical input parameters, the returned \texttt{jitter} and \texttt{total} values may differ across executions.
This dependency on randomness introduces non-deterministic behavior into the API.
For example, the test shown in Figure~\ref{fig:flaky-test} interacting with this endpoint may pass in some executions and fail in others, even sending the same request.


\begin{figure}
\begin{lstlisting}[numbers=none,language=Kotlin,basicstyle=\scriptsize]
@GetMapping("/price/estimate")
fun estimatePrice(@RequestParam base: Int): Map<String, Int> {
	val randomJitter = randomInt(10) 
	
	val total = base + randomJitter
	
	return mapOf(
		"base" to base,
		"jitter" to randomJitter,
		"total" to total
	)
}
\end{lstlisting}
	\squeezeupfigureAbove
\caption{\label{fig:flaky-example}
	Snippet of an example of an endpoint with flakiness
}
\squeezeupfigureAbove
\end{figure}

\begin{figure}
\begin{lstlisting}[numbers=none,language=Kotlin,basicstyle=\scriptsize]
@Test @Timeout(60)
fun test_8_getOnEstimateReturnsObject()  {
	
	given().accept("*/*")
		.header("x-EMextraHeader123", "")
		.get("${baseUrlOfSut}/api/flakinessdetect/price/estimate?" + 
				"base=666&" + 
				"EMextraParam123=_EM_2_XYZ_")
		.then()
		.statusCode(200)
		.assertThat()
		.contentType("application/json")
		.body("'base'", numberMatches(666))
		.body("'jitter'", numberMatches(3)) // may vary due to randomness
		.body("'total'", numberMatches(669)) // depends on the generated jitter
}
\end{lstlisting}
	\caption{\label{fig:flaky-test}
		Snippet of an example test linked to flaky endpoints
	}
\end{figure}

\squeezeup
\subsection{EvoMaster}

\evo is an open-source evolutionary fuzzer for automated REST API testing, which has been actively maintained since its inception in 2016~\cite{arcuri2017restful}.
It is one of the few fuzzers that support both black-box and white-box testing of REST APIs, enabling test generation with and without access to the source code.
Its white-box mode is built on instrumentation that is developed to automatically collect runtime execution information, such as statement coverage and executed SQL commands.
Based on this runtime feedback, \evo employs an evolutionary algorithm: Many Independent Objective (MIO) algorithm~\cite{arcuri2018test}, which is specifically designed for system-level testing to effectively handle many and potentially conflicting testing objectives, e.g., \evo treats each predicate, branch, and line of code as a separate testing objective.
In its black-box mode, \evo adopts random strategies by default to guide test generation based on API schema (e.g., OpenAPI specifications for REST APIs) and request–response coverage.

Over the years, \evo has been integrated with various advanced techniques, e.g., testability transformation~\cite{arcuri2021enhancing}, SQL handling~\cite{arcuri2020sql,arcuri2023advanced}, resource-based strategies~\cite{zhang2021resource}, adaptive hypermutation~\cite{zhang2021adaptive}, mocking~\cite{seran2025handling}, and security testing~\cite{arcuri2025fuzzing}.
SQL handling and mocking for managing web service interactions can help mitigate flakiness in Web APIs.
Beyond REST APIs, \evo also supports fuzzing of GraphQL APIs~\cite{belhadi2023random} and RPC-based APIs~\cite{zhang2023rpc,zhang2024seeding,zhang2025fuzzing}.
Empirical studies on REST API fuzzers applied to open-source REST APIs show that \evo in white-box mode achieves state-of-the-art performance in terms of code coverage and fault detection~\cite{Kim2022Rest,zhang2023open}.
Moreover, its effectiveness in both black-box and white-box modes has been demonstrated in industrial settings, including large-scale enterprise systems at companies such as Volkswagen~\cite{icst2025vw,poth2025technology} and Meituan~\cite{zhang2023rpc,zhang2025fuzzing}.

\section{Related Work}
\label{sec:relatedwork}

In this section, we discuss the related work from three aspects: API fuzzing, flakiness at the unit-testing and system testing levels. 

\subsection{API Fuzzing}
\label{subsec:fuzzers}

API Fuzzing (or API Fuzz Testing, API Testing) has received significant attention in both academia and industry.
In the literature, most approaches target black-box testing, often relying on API specification (e.g., OpenAPI) or analyzing requests and responses.
In contract, few studies have explored white-box testing by utilizing runtime execution information to guide test generation.  

{
\sloppy
In API testing, REST APIs remain the dominant research problem, likely due to their early introduction and widespread adoption in modern web services~\cite{golmohammadi2023testing,Kim2022Rest,zhang2023open}.
Researchers have proposed various techniques for REST API testing, e.g.,
\apif~\cite{wang2024beyond},
\apirl~\cite{foley2025apirl},
\aratrl~\cite{kim2023adaptive},
\astra~\cite{sondhi2025utilizing},
\autoresttest~\cite{kim2025autoresttest},
\bboxrt~\cite{laranjeiro2021black},
\deeprest~\cite{corradini2024deeprest},
\emrest~\cite{xu2025effective},
\evo~\cite{arcuri2025tool},
\kat~\cite{le2024kat},
\llamaresttest~\cite{kim2025llamaresttest},
\logiaagent~\cite{zhang2025logiagent},
\miner~\cite{lyu2023miner},
\morest~\cite{liu2022icse},
\nautilus~\cite{deng2023nautilus},
\openapifuzzer~\cite{ferech2023efficient},
\raft~\cite{saha2025rest},
\restct~\cite{wu2022icse},
\restest~\cite{martinLopez2021Restest},
\restler~\cite{restlerICSE2019},
\resttestgen~\cite{viglianisi2020resttestgen},
\schemathesis~\cite{hatfield2022deriving},
\voapi~\cite{du2024vulnerability}
and
\wuppiefuzz~\cite{rooijakkers2025wuppiefuzz}.
For example, \restler employs dynamic analysis of request–response dependencies to guide test generation. 
\morest and \resttestgen constructs model or graph for driving test generation, while \schemathesis is developed in the context of property-based testing.
Recent works such as
\apif~\cite{wang2024beyond},
\apirl~\cite{foley2025apirl},
\aratrl~\cite{kim2023adaptive},
\astra~\cite{sondhi2025utilizing},
\deeprest~\cite{corradini2024deeprest},
\kat~\cite{le2024kat},
\llamaresttest~\cite{kim2025llamaresttest}, and
\logiaagent~\cite{zhang2025logiagent} 
leverage artificial intelligence techniques to improve test effectiveness. 
\evo~\cite{arcuri2025tool} is the approach that supports both black-box and white-box testing.

GraphQL API testing has also been attracting attention~\cite{quina2023graphql}. 
GraphQL APIs, which allow clients to flexibly query structured data, introduce new testing challenges due to their complex query schema and nested request structures. 
Recent work has proposed using search-based testing techniques~\cite{belhadi2023random}, property-based testing~\cite{karlsson2020automatic}, and mutation-based testing~\cite{pan2025trailblazer} for GraphQL APIs.

Remote Procedure Call (RPC) is widely adopted in industry for high-performance communication in microservice architectures~\cite{zhang2025fuzzing,zhang2024seeding,zhang2023rpc,liu2022record}. 
Testing RPC-based APIs is challenging due to the diversity of frameworks (e.g., gRPC, Thrift, and Dubbo) and their complex inter-service dependencies. 
Existing studies have proposed various techniques to address these challenges, including white-box fuzzing~\cite{zhang2023rpc,zhang2025fuzzing}, seeding and mocking~\cite{zhang2024seeding}, traffic recording and replay~\cite{liu2022record}, and Protobuf-schema based testing approaches~\cite{wang2023zero}.
}

\subsection{Flakiness at the Unit Testing Level}
In the literature, there is a rich body of research that investigates the challenge of flaky unit tests across various programming languages, including Java, Python, and .NET. and a comprehensive pipeline for the detection, classification, prediction, and mitigation of flaky unit tests has been established by leveraging techniques ranging from traditional static and dynamic analyses to advanced AI techniques such as LLMs. 
A comprehensive survey~\cite{parry2021survey} has been conducted to summarize the existing body of the literature on flaky tests, hence in this section we mainly focus on the most recent studies within the last few years since then.

Leesatapornwongsa et al.~\cite{leesatapornwongsa2022flakerepro} proposed \textit{FlakeRepro} to help developers reproduce failed executions of flaky tests caused by concurrency, by combining both static and dynamic analysis. 
Akli et al.~\cite{akli2023flakycat} proposed \textit{FlakyCat}, a method for predicting flaky test categories by relying on CodeBERT. 
Fatima et al.~\cite{fatima2024flakyfix} proposed \textit{FlakyFix}, a method for predicting a fix category (out of 13 in total) for a flaky test by analyzing the test code only with CodeBERT and UniXcoder, and leverages these labels to guide LLMs in generating successful automated repairs. Along the same line, Rahman and Shi proposed \textit{FlakeSync}~\cite{rahman2024flakesync} to repair async flaky tests by introducing synchronization for a specific test execution.

Gruber et al.~\cite{gruber2024automatic} conducted a comprehensive empirical study by analyzing over 6,000 Java and Python projects, on which each test generated by two test generation tools was executed 200 times. Results reveal that: 1) automated test generation tools produce flaky tests even more frequently than developers do; 2) generated flaky tests are often caused by randomness and unspecified behaviours, while developers write flaky tests more often caused by concurrency and networking operations; and 3) flakiness suppression mechanisms are effective in reducing the number of tests but also revealing previously-unknown types of flakiness. 

Li et al.~\cite{li2025hiflaky} proposed \textit{HiFlaky}, a method that detects and classify flaky tests by considering hierarchical dependencies among the root causes of flakiness in test code
Their empirical study results show that HiFlaky achieves higher flaky test detection accuracy than two baselines: \textit{FlakeFlagger}~\cite{alshammari2021flakeflagger} and \textit{Fakify}~\cite{fatima2022flakify}, which are both flaky test predictors. 
Moreover, when observing the increase use of LLMs for classifying flaky tests, Rahman et al.~\cite{rahman2024quantizing} proposed \textit{FlakyQ}, a framework that utilizes quantized LLMs for feature extraction coupled with traditional machine learning classifiers to achieve the reduction in prediction time and memory usage without compromising accuracy. 
Parry et al.~\cite{parry2025test} introduced the concept of \textit{FLIMsiness}, a form of non-determinism in unit tests that is intentionally induced by applying mutation operators to the code under test, and demonstrates that filmsiness can significantly detect more flaky tests than traditional rerunning strategies. 
Schroeder et al.~\cite{schroeder2025preliminary} conducted a preliminary study of fixed flaky tests in Rust projects and identified nine common root causes of test flakiness. 

Note that the above-mentioned body of work focuses specifically on unit testing to address non-determinism through automated detection, root cause categorization, and code repair.
Our work focuses on system testing, where tests are composed of sequences of HTTP calls towards the tested API.
Assertions are based on what returned in these calls, e.g., HTTP headers and body payloads.
Calls over a network, and interactions with databases, might introduce further sources of flakiness not commonly seen in unit testing.

\squeezeup
\subsection{Flakiness at the System Testing Level}
While a rich body of research has established for managing flakiness at the unit testing level, studies focusing on system testing remain comparatively scarce. In the rest of the section, we discuss some of these works that investigate flakiness in complex domains such as database management systems and autonomous driving simulators.

Mor{\'a}n et al.~\cite{moran2019debugging} proposed \textit{FlakcLoc} , a spectrum-based localization technique to identify the root causes of flakiness in web applications by analyzing how uncontrolled environmental factors (e.g., network latency, memory) trigger inconsistent test outputs.
Dong et al.~\cite{dong2020concurrency} proposed \textit{FlakeScanner} to detect flaky tests for Android apps by systematically exploring event orders, along with a benchmark named \textit{FlakyAppRepo} for enabling the study of GUI test flakiness. 
Ngo et al.~\cite{ngo2022research} provided a review of how academia and industry handle test flakiness and highlighted that current research trends indicate a concentration on unit-level flakiness and leaves end-to-end system testing comparatively under-explored. 
Osikowicz et al.~\cite{osikowicz2025empirically} recently conducted an empirical study to investigate flaky tests in simulation based autonomous driving testing, and observed that one-third of driving scenarios executed in CARLA are potentially flaky due to unintentional nondeterminism in simulators (e.g., bugs and the use of rendering engine). 
Berndt et al.~\cite{berndt2026flakiness} recently conducted a study to investigate the flakiness of LLM-generated tests for database management systems. Results of their study found that LLM-generated tests are often more prone to flakiness than original test suites, largely due to non-deterministic data ordering, and the flakiness transfer (from existing tests to the newly generated ones via prompt) is more prevalent in closed-source database systems than in open-source ones. 

To our knowledge, there is limited work that systematically studies and addresses flakiness in the context of web services such as REST APIs, despite the substantial focus on REST API fuzzing in prior work. In contrast, our approach \method specifically targets the detection and mitigation of flakiness in REST API fuzzing.

\section{Empirical Analysis on Flakiness of REST APIs}\label{sec:empiricalStudy}

To better understand flakiness existing in REST APIs, we carried out a comprehensive empirical analysis to answer these two questions:

\begin{description}
	\item[{\bf RQ1}:] \rqA
	\item[{\bf RQ2}:] \rqB
\end{description}

\subsection{Experiment Settings and Design}

\textbf{Open-Source REST APIs.}
Several benchmarks exist for evaluating fuzzing techniques (e.g.,~\cite{hazimeh2020magma,li2021unifuzz,metzman2021fuzzbench,bohme2022reliability,ounjai2023green,miao2025program}).
In our case, as we focus on REST APIs,
we selected Web Fuzzing Dataset (WFD)~\cite{sahin2025wfc}, previously known as EMB~\cite{icst2023emb}.
This dataset is the most used in the research literature of fuzzing REST APIs, extended each year with new APIs since 2017.
With 36 APIs, it currently provides the largest publicly available collection of REST APIs with source code and experiment infrastructure.
Table~\ref{tab:sutinfo} shows these 36 REST APIs along with their descriptive statistics, e.g., the number of source files (\#Files), lines of code (LoCs), endpoints (\#End.), runtime environments,
and databases used. 
These APIs cover a broad range of characteristics, e.g., LoCs ranging from 117 to 174,781, 1 to 258 endpoints, from with no database usage to with multiple connections across different databases.
Overall, our analysis covers 36 REST APIs comprising 6,465 source files, 657,162 LoCs, and 1,487 endpoints, with 25 out of 36 APIs interacting with databases.

\begin{table}
	\centering
	\small
	\caption{Descriptive information of the REST APIs employed}
	\label{tab:sutinfo}
	\squeezeupfigureAbove
		\begin{tabular}{l rrr l p{2cm}}\\ 
\toprule 
SUT  &\#Files & \#LOCs & \#End. & Runtime & Databases\\ 
\midrule 
\emph{bibliothek} &  33 &  2176 &  8 & JDK 17 & MongoDB \\ 
\emph{blogapi} &  89 &  4787 &  52 & JDK 8 & MySQL \\ 
\emph{catwatch} &  106 &  9636 &  14 & JDK 8 & H2 \\ 
\emph{cwa-verification} &  47 &  3955 &  5 & JDK 11 & H2 \\ 
\emph{erc20-rest-service} &  7 &  1378 &  13 & JDK 8 &  \\ 
\emph{familie-ba-sak} &  1089 &  143556 &  183 & JDK 17 & PostgreSQL \\ 
\emph{features-service} &  39 &  2275 &  18 & JDK 8 & H2 \\ 
\emph{genome-nexus} &  405 &  30004 &  23 & JDK 8 & MongoDB \\ 
\emph{gestaohospital} &  33 &  3506 &  20 & JDK 8 & MongoDB \\ 
\emph{http-patch-spring} &  30 &  1450 &  6 & JDK 11 &  \\ 
\emph{languagetool} &  1385 &  174781 &  2 & JDK 8 &  \\ 
\emph{market} &  124 &  9861 &  13 & JDK 11 & H2 \\ 
\emph{microcks} &  471 &  66186 &  88 & JDK 21 & MongoDB \\ 
\emph{ocvn} &  526 &  45521 &  258 & JDK 8 & H2, MongoDB \\ 
\emph{ohsome-api} &  87 &  14166 &  134 & JDK 17 & OSHDB \\ 
\emph{pay-publicapi} &  377 &  34576 &  10 & JDK 11 & Redis \\ 
\emph{person-controller} &  16 &  1112 &  12 & JDK 21 & MongoDB \\ 
\emph{proxyprint} &  73 &  8338 &  74 & JDK 8 & H2 \\ 
\emph{quartz-manager} &  129 &  5068 &  11 & JDK 11 &  \\ 
\emph{reservations-api} &  39 &  1853 &  7 & JDK 11 & MongoDB \\ 
\emph{rest-ncs} &  9 &  605 &  6 & JDK 8 &  \\ 
\emph{rest-news} &  11 &  857 &  7 & JDK 8 & H2 \\ 
\emph{rest-scs} &  13 &  862 &  11 & JDK 8 &  \\ 
\emph{restcountries} &  24 &  1977 &  22 & JDK 8 &  \\ 
\emph{scout-api} &  93 &  9736 &  49 & JDK 8 & H2 \\ 
\emph{session-service} &  15 &  1471 &  8 & JDK 8 & MongoDB \\ 
\emph{spring-actuator-demo} &  5 &  117 &  2 & JDK 8 &  \\ 
\emph{spring-batch-rest} &  65 &  3668 &  5 & JDK 8 &  \\ 
\emph{spring-ecommerce} &  58 &  2223 &  26 & JDK 8 & MongoOB, Redis, Elasticsearch \\ 
\emph{spring-rest-example} &  32 &  1426 &  9 & JDK 17 & MySQL \\ 
\emph{swagger-petstore} &  23 &  1631 &  19 & JDK 8 &  \\ 
\emph{tiltaksgjennomforing} &  472 &  27316 &  79 & JDK 17 & PostgreSQL \\ 
\emph{tracking-system} &  87 &  5947 &  67 & JDK 11 & H2 \\ 
\emph{user-management} &  69 &  4274 &  21 & JDK 8 & MySQL \\ 
\emph{webgoat} &  355 &  27638 &  204 & JDK 21 & H2 \\ 
\emph{youtube-mock} &  29 &  3229 &  1 & JDK 8 &  \\ 
\midrule 
Total 36 & 6465 & 657162 & 1487 & &  25 \\ 
\bottomrule 
\end{tabular} 

	\squeezeupfigureAfter
\end{table}

\textbf{Fuzzer Selection.}
Since test flakiness can only be identified through repeated executions of test cases, its analysis requires executable tests that exercise diverse inputs, request sequences, and state-dependent interactions, in order to increase the likelihood of exposing non-deterministic behavior.
Fuzzers that can automatically generate such tests are well suited for our flakiness analysis.
Moreover, flakiness characteristics of tests generated by white-box and black-box fuzzers may differ. 
For example, tests generated by white-box approaches may manipulate internal system states, including external web interactions~\cite{seran2025handling}, as well as SQL~\cite{arcuri2020sql,arcuri2024advanced} and NoSQL databases~\cite{ghianni2025search}, which may either introduce or reduce flakiness.
Moreover, empirical studies involving both black-box and white-box fuzzers have also shown that \evo, when used in white-box mode, achieves the best performance in terms of code coverage and fault detection~\cite{Kim2022Rest,zhang2023open}. 
Hence, we selected \evo because it supports both testing modes and enables the analysis of flakiness in tests generated by different fuzzing techniques.


%

\textbf{Experiment Settings.}
We applied \evo in both black-box (denoted as BB) and white-box mode (denoted as WB) to each of the 36 APIs, using a one-hour search budget, which is the most commonly adopted configuration in the REST API fuzzing literature~\cite{Kim2022Rest,zhang2023open}. 
Considering the randomness of the fuzzer, we run the test generations 10 times for each configuration.
As the runtime environment may also affect test behavior, we executed generated test cases under two execution environments:
\begin{itemize}
	\item \fuzzenv: a DELL Precision 7875 Tower equipped with an AMD Ryzen Threadripper PRO 7975WX (64 cores, up to 5.35~GHz), 128~GB of RAM, running 64-bit Ubuntu 22.04.5;
	\item \execenv: a ThinkStation P620 equipped with an AMD Ryzen Threadripper PRO 5995WX (128 cores, up to 4.58~GHz), 256G~GB of RAM, running 64-bit Ubuntu 24.04.3.
\end{itemize}
To observe the non-deterministic behavior of these test cases, we compiled and executed tests generated by each configuration 100 times in both \fuzzenv and \execenv.

\subsection{Results of Flaky Tests}
\label{subsec:rq1results}

Table~\ref{tab:rq1_false_flakytests} reports, for each API on each environment, the average failure rate ($FR$\%) and the standard error ($sd_r$) across the 10 runs for both BB and WB modes of \evo.
To characterize the stability of flakiness, we also report the numbers of failed tests (\#$F$), consistently failed tests  (\#$F_c$), and unstable failed tests  (\#$F_u$).
Overall, we can observe that across the two execution environments, with two generation strategies, among 36 APIs, flaky tests were observed for 31 APIs in both \fuzzenv and \execenv, indicating that flakiness is common rather than an exception in REST APIs.

\begin{table}
	\centering
	\small
	\caption{Results of the number of generated tests (\#$T$), the failure rate ($FR$\%), the number of failed tests (\#$F$), the number of consistent failed tests (\#$F_c$), and the number of unstable failed tests (\#$F_u$) on the \fuzzenv and \execenv environments.}
	\label{tab:rq1_false_flakytests}
	\squeezeupfigureAbove
	\resizebox{.92\linewidth}{!}{
		\begin{tabular}{ll r r | rr | rr}\\ 
\toprule 
 & & & & & \fuzzenv & & \execenv \\ 
SUT & Mode & \#$T$ & Lines\% & $FR$\% ($sd_r$) & \#$F$ (\#$F_c$, \#$F_u$)  & $FR$\% ($sd_r$) & \#$F$ (\#$F_c$, \#$F_u$)  \\ 
\midrule 
\emph{\bibliothek} & \cellcolor{gray!20}BB & \cellcolor{gray!20}10.0 & \cellcolor{gray!20}27.1 & \cellcolor{gray!20}0.0 (0.0) & \cellcolor{gray!20}0.0 (0.0, 0.0) & \cellcolor{gray!20}0.0 (0.0) & \cellcolor{gray!20}0.0 (0.0, 0.0)\\  
 & WB & 13.4 & 26.7 & 0.0 (0.0) & 0.0 (0.0, 0.0) & 0.0 (0.0) & 0.0 (0.0, 0.0)\\  
\emph{\blogapi} & \cellcolor{gray!20}BB & \cellcolor{gray!20}178.1 & \cellcolor{gray!20}24.6 & \cellcolor{gray!20}9.0 (0.2) & \cellcolor{gray!20}16.0 (16.0, 0.0) & \cellcolor{gray!20}9.0 (0.2) & \cellcolor{gray!20}16.0 (16.0, 0.0)\\  
 & WB & 201.2 & 31.5 & $\textcolor{blue}{\triangledown}$3.4 (0.8) & $\textcolor{blue}{\triangledown}$6.9 ($\textcolor{blue}{\triangledown}$6.9, 0.0) & $\textcolor{red}{\triangle}$10.4 (1.5) & $\textcolor{red}{\triangle}$21.0 ($\textcolor{red}{\triangle}$21.0, 0.0)\\  
\emph{\catwatch} & \cellcolor{gray!20}BB & \cellcolor{gray!20}56.1 & \cellcolor{gray!20}29.7 & \cellcolor{gray!20}7.1 (2.3) & \cellcolor{gray!20}4.0 (2.0, 2.0) & \cellcolor{gray!20}7.1 (2.4) & \cellcolor{gray!20}4.0 (2.2, 1.8)\\  
 & WB & 115.5 & 44.8 & $\textcolor{blue}{\triangledown}$4.7 (2.4) & $\textcolor{red}{\triangle}$5.4 ($\textcolor{red}{\triangle}$5.4, $\textcolor{blue}{\blacktriangledown}$0.0) & $\textcolor{red}{\triangle}$9.4 (16.1) & $\textcolor{red}{\triangle}$11.1 ($\textcolor{red}{\triangle}$5.7, $\textcolor{red}{\triangle}$5.4)\\  
\emph{\cwaverification} & \cellcolor{gray!20}BB & \cellcolor{gray!20}5.0 & \cellcolor{gray!20}37.9 & \cellcolor{gray!20}100.0 (0.0) & \cellcolor{gray!20}5.0 (5.0, 0.0) & \cellcolor{gray!20}100.0 (0.0) & \cellcolor{gray!20}5.0 (5.0, 0.0)\\  
 & WB & 5.0 & 16.7 & 100.0 (0.0) & 5.0 (5.0, 0.0) & 100.0 (0.0) & 5.0 (5.0, 0.0)\\  
\emph{\ercrestservice} & \cellcolor{gray!20}BB & \cellcolor{gray!20}13.0 & \cellcolor{gray!20}25.4 & \cellcolor{gray!20}0.0 (0.0) & \cellcolor{gray!20}0.0 (0.0, 0.0) & \cellcolor{gray!20}0.0 (0.0) & \cellcolor{gray!20}0.0 (0.0, 0.0)\\  
 & WB & 20.4 & 31.2 & 0.0 (0.0) & 0.0 (0.0, 0.0) & 0.0 (0.0) & 0.0 (0.0, 0.0)\\  
\emph{\familiebasak} & \cellcolor{gray!20}BB & \cellcolor{gray!20}1326.5 & \cellcolor{gray!20}14.4 & \cellcolor{gray!20}0.6 (0.1) & \cellcolor{gray!20}8.4 (8.4, 0.0) & \cellcolor{gray!20}3.1 (7.8) & \cellcolor{gray!20}41.9 (8.4, 33.5)\\  
 & WB & 187.6 & 17.5 & $\textcolor{red}{\triangle}$1.1 (0.0) & $\textcolor{blue}{\triangledown}$2.0 ($\textcolor{blue}{\triangledown}$2.0, 0.0) & $\textcolor{red}{\triangle}$8.9 (1.6) & $\textcolor{blue}{\triangledown}$17.0 ($\textcolor{red}{\triangle}$17.0, $\textcolor{blue}{\blacktriangledown}$0.0)\\  
\emph{\featuresservice} & \cellcolor{gray!20}BB & \cellcolor{gray!20}26.3 & \cellcolor{gray!20}57.0 & \cellcolor{gray!20}4.9 (3.3) & \cellcolor{gray!20}1.3 (1.0, 0.3) & \cellcolor{gray!20}4.9 (3.3) & \cellcolor{gray!20}1.3 (1.0, 0.3)\\  
 & WB & 17.8 & 39.9 & $\textcolor{blue}{\triangledown}$0.6 (1.8) & $\textcolor{blue}{\triangledown}$0.1 ($\textcolor{blue}{\triangledown}$0.1, $\textcolor{blue}{\blacktriangledown}$0.0) & $\textcolor{blue}{\triangledown}$0.6 (1.8) & $\textcolor{blue}{\triangledown}$0.1 ($\textcolor{blue}{\triangledown}$0.1, $\textcolor{blue}{\blacktriangledown}$0.0)\\  
\emph{\genomenexus} & \cellcolor{gray!20}BB & \cellcolor{gray!20}43.9 & \cellcolor{gray!20}28.1 & \cellcolor{gray!20}28.2 (2.9) & \cellcolor{gray!20}12.4 (7.8, 4.6) & \cellcolor{gray!20}20.9 (1.1) & \cellcolor{gray!20}9.2 (8.8, 0.4)\\  
 & WB & 76.3 & 35.8 & $\textcolor{blue}{\triangledown}$23.2 (6.3) & $\textcolor{red}{\triangle}$18.1 ($\textcolor{red}{\triangle}$17.4, $\textcolor{blue}{\triangledown}$0.7) & $\textcolor{red}{\triangle}$22.6 (6.5) & $\textcolor{red}{\triangle}$17.7 ($\textcolor{red}{\triangle}$17.7, $\textcolor{blue}{\blacktriangledown}$0.0)\\  
\emph{\gestaohospital} & \cellcolor{gray!20}BB & \cellcolor{gray!20}38.6 & \cellcolor{gray!20}42.5 & \cellcolor{gray!20}37.5 (9.6) & \cellcolor{gray!20}15.0 (14.5, 0.5) & \cellcolor{gray!20}37.5 (9.6) & \cellcolor{gray!20}15.0 (14.5, 0.5)\\  
 & WB & 121.0 & 43.0 & $\textcolor{blue}{\triangledown}$7.2 (4.3) & $\textcolor{blue}{\triangledown}$8.7 ($\textcolor{blue}{\triangledown}$6.6, $\textcolor{red}{\triangle}$2.1) & $\textcolor{blue}{\triangledown}$7.6 (4.0) & $\textcolor{blue}{\triangledown}$9.2 ($\textcolor{blue}{\triangledown}$8.7, 0.5)\\  
\emph{\httppatchspring} & \cellcolor{gray!20}BB & \cellcolor{gray!20}22.8 & \cellcolor{gray!20}55.1 & \cellcolor{gray!20}25.4 (3.4) & \cellcolor{gray!20}5.8 (5.0, 0.8) & \cellcolor{gray!20}25.4 (3.4) & \cellcolor{gray!20}5.8 (5.0, 0.8)\\  
 & WB & 26.6 & 57.9 & $\textcolor{red}{\triangle}$38.1 (5.1) & $\textcolor{red}{\triangle}$10.2 ($\textcolor{red}{\triangle}$10.2, $\textcolor{blue}{\blacktriangledown}$0.0) & $\textcolor{red}{\triangle}$46.1 (4.9) & $\textcolor{red}{\triangle}$12.3 ($\textcolor{red}{\triangle}$12.3, $\textcolor{blue}{\blacktriangledown}$0.0)\\  
\emph{\languagetool} & \cellcolor{gray!20}BB & \cellcolor{gray!20}3.9 & \cellcolor{gray!20}9.0 & \cellcolor{gray!20}0.0 (0.0) & \cellcolor{gray!20}0.0 (0.0, 0.0) & \cellcolor{gray!20}0.0 (0.0) & \cellcolor{gray!20}0.0 (0.0, 0.0)\\  
 & WB & 37.8 & 26.6 & $\textcolor{red}{\blacktriangle}$0.2 (0.6) & $\textcolor{red}{\blacktriangle}$0.1 (0.0, $\textcolor{red}{\blacktriangle}$0.1) & $\textcolor{red}{\blacktriangle}$1.0 (2.0) & $\textcolor{red}{\blacktriangle}$0.5 ($\textcolor{red}{\blacktriangle}$0.1, $\textcolor{red}{\blacktriangle}$0.4)\\  
\emph{\market} & \cellcolor{gray!20}BB & \cellcolor{gray!20}53.5 & \cellcolor{gray!20}25.9 & \cellcolor{gray!20}21.4 (3.8) & \cellcolor{gray!20}11.5 (5.5, 6.0) & \cellcolor{gray!20}21.4 (3.8) & \cellcolor{gray!20}11.5 (5.6, 5.9)\\  
 & WB & 43.9 & 44.8 & $\textcolor{blue}{\triangledown}$21.0 (3.1) & $\textcolor{blue}{\triangledown}$9.2 ($\textcolor{blue}{\triangledown}$5.1, $\textcolor{blue}{\triangledown}$4.1) & $\textcolor{red}{\triangle}$36.1 (4.2) & $\textcolor{red}{\triangle}$15.7 ($\textcolor{red}{\triangle}$11.6, $\textcolor{blue}{\triangledown}$4.1)\\  
\emph{\microcks} & \cellcolor{gray!20}BB & \cellcolor{gray!20}289.2 & \cellcolor{gray!20}11.4 & \cellcolor{gray!20}13.7 (1.1) & \cellcolor{gray!20}39.7 (36.4, 3.3) & \cellcolor{gray!20}14.8 (1.1) & \cellcolor{gray!20}42.7 (39.4, 3.3)\\  
 & WB & 94.4 & 23.0 & $\textcolor{blue}{\triangledown}$1.5 (0.5) & $\textcolor{blue}{\triangledown}$1.4 ($\textcolor{blue}{\triangledown}$1.4, $\textcolor{blue}{\blacktriangledown}$0.0) & $\textcolor{blue}{\triangledown}$2.5 (0.5) & $\textcolor{blue}{\triangledown}$2.4 ($\textcolor{blue}{\triangledown}$2.4, $\textcolor{blue}{\blacktriangledown}$0.0)\\  
\emph{\ocvn} & \cellcolor{gray!20}BB & \cellcolor{gray!20}2045.1 & \cellcolor{gray!20}19.8 & \cellcolor{gray!20}28.6 (0.5) & \cellcolor{gray!20}584.7 (96.2, 488.5) & \cellcolor{gray!20}28.6 (0.5) & \cellcolor{gray!20}584.8 (97.2, 487.6)\\  
 & WB & 1315.5 & 21.0 & $\textcolor{red}{\triangle}$55.2 (1.5) & $\textcolor{red}{\triangle}$725.2 ($\textcolor{blue}{\triangledown}$56.6, $\textcolor{red}{\triangle}$668.6) & $\textcolor{red}{\triangle}$67.3 (0.8) & $\textcolor{red}{\triangle}$887.3 ($\textcolor{red}{\triangle}$887.3, $\textcolor{blue}{\blacktriangledown}$0.0)\\  
\emph{\ohsomeapi} & \cellcolor{gray!20}BB & \cellcolor{gray!20}627.8 & \cellcolor{gray!20}41.2 & \cellcolor{gray!20}0.0 (0.0) & \cellcolor{gray!20}0.0 (0.0, 0.0) & \cellcolor{gray!20}0.0 (0.0) & \cellcolor{gray!20}0.0 (0.0, 0.0)\\  
 & WB & 640.0 & 23.9 & $\textcolor{red}{\blacktriangle}$0.1 (0.3) & $\textcolor{red}{\blacktriangle}$0.8 ($\textcolor{red}{\blacktriangle}$0.8, 0.0) & $\textcolor{red}{\blacktriangle}$0.1 (0.3) & $\textcolor{red}{\blacktriangle}$0.8 ($\textcolor{red}{\blacktriangle}$0.8, 0.0)\\  
\emph{\paypublicapi} & \cellcolor{gray!20}BB & \cellcolor{gray!20}60.4 & \cellcolor{gray!20}15.3 & \cellcolor{gray!20}66.7 (6.2) & \cellcolor{gray!20}40.5 (0.0, 40.5) & \cellcolor{gray!20}0.0 (0.0) & \cellcolor{gray!20}0.0 (0.0, 0.0)\\  
 & WB & 38.4 & 13.0 & $\textcolor{blue}{\triangledown}$1.8 (1.7) & $\textcolor{blue}{\triangledown}$0.7 ($\textcolor{red}{\blacktriangle}$0.7, $\textcolor{blue}{\blacktriangledown}$0.0) & $\textcolor{red}{\blacktriangle}$1.8 (1.7) & $\textcolor{red}{\blacktriangle}$0.7 ($\textcolor{red}{\blacktriangle}$0.7, 0.0)\\  
\emph{\personcontroller} & \cellcolor{gray!20}BB & \cellcolor{gray!20}14.0 & \cellcolor{gray!20}54.0 & \cellcolor{gray!20}34.3 (6.6) & \cellcolor{gray!20}4.8 (4.8, 0.0) & \cellcolor{gray!20}34.3 (6.6) & \cellcolor{gray!20}4.8 (4.8, 0.0)\\  
 & WB & 32.3 & 62.0 & $\textcolor{blue}{\triangledown}$29.9 (3.8) & $\textcolor{red}{\triangle}$9.6 ($\textcolor{red}{\triangle}$9.6, 0.0) & $\textcolor{blue}{\triangledown}$29.9 (3.8) & $\textcolor{red}{\triangle}$9.6 ($\textcolor{red}{\triangle}$9.6, 0.0)\\  
\emph{\proxyprint} & \cellcolor{gray!20}BB & \cellcolor{gray!20}556.6 & \cellcolor{gray!20}6.8 & \cellcolor{gray!20}83.8 (0.2) & \cellcolor{gray!20}466.5 (466.5, 0.0) & \cellcolor{gray!20}83.8 (0.2) & \cellcolor{gray!20}466.7 (466.7, 0.0)\\  
 & WB & 380.4 & 30.3 & $\textcolor{blue}{\triangledown}$2.9 (1.5) & $\textcolor{blue}{\triangledown}$11.1 ($\textcolor{blue}{\triangledown}$10.7, $\textcolor{red}{\blacktriangle}$0.4) & $\textcolor{blue}{\triangledown}$45.4 (1.2) & $\textcolor{blue}{\triangledown}$172.6 ($\textcolor{blue}{\triangledown}$172.3, $\textcolor{red}{\blacktriangle}$0.3)\\  
\emph{\quartzmanager} & \cellcolor{gray!20}BB & \cellcolor{gray!20}27.0 & \cellcolor{gray!20}28.2 & \cellcolor{gray!20}29.6 (0.0) & \cellcolor{gray!20}8.0 (2.0, 6.0) & \cellcolor{gray!20}29.6 (0.0) & \cellcolor{gray!20}8.0 (2.0, 6.0)\\  
 & WB & 50.8 & 37.7 & $\textcolor{blue}{\triangledown}$13.1 (6.7) & $\textcolor{blue}{\triangledown}$6.6 ($\textcolor{red}{\triangle}$6.6, $\textcolor{blue}{\blacktriangledown}$0.0) & $\textcolor{blue}{\triangledown}$13.1 (6.7) & $\textcolor{blue}{\triangledown}$6.6 ($\textcolor{red}{\triangle}$6.6, $\textcolor{blue}{\blacktriangledown}$0.0)\\  
\emph{\reservationsapi} & \cellcolor{gray!20}BB & \cellcolor{gray!20}23.0 & \cellcolor{gray!20}44.8 & \cellcolor{gray!20}12.6 (1.4) & \cellcolor{gray!20}2.9 (2.9, 0.0) & \cellcolor{gray!20}12.6 (1.4) & \cellcolor{gray!20}2.9 (2.9, 0.0)\\  
 & WB & 44.1 & 54.0 & $\textcolor{red}{\triangle}$37.9 (42.2) & $\textcolor{red}{\triangle}$16.1 ($\textcolor{red}{\triangle}$4.8, $\textcolor{red}{\blacktriangle}$11.3) & $\textcolor{red}{\triangle}$63.8 (46.8) & $\textcolor{red}{\triangle}$27.6 ($\textcolor{red}{\triangle}$4.8, $\textcolor{red}{\blacktriangle}$22.8)\\  
\emph{\restncs} & \cellcolor{gray!20}BB & \cellcolor{gray!20}22.1 & \cellcolor{gray!20}60.5 & \cellcolor{gray!20}0.0 (0.0) & \cellcolor{gray!20}0.0 (0.0, 0.0) & \cellcolor{gray!20}0.0 (0.0) & \cellcolor{gray!20}0.0 (0.0, 0.0)\\  
 & WB & 40.3 & 79.9 & 0.0 (0.0) & 0.0 (0.0, 0.0) & 0.0 (0.0) & 0.0 (0.0, 0.0)\\  
\emph{\restnews} & \cellcolor{gray!20}BB & \cellcolor{gray!20}18.1 & \cellcolor{gray!20}47.5 & \cellcolor{gray!20}40.3 (7.6) & \cellcolor{gray!20}7.3 (5.9, 1.4) & \cellcolor{gray!20}40.3 (7.6) & \cellcolor{gray!20}7.3 (5.9, 1.4)\\  
 & WB & 45.8 & 61.8 & $\textcolor{blue}{\triangledown}$3.7 (1.7) & $\textcolor{blue}{\triangledown}$1.7 ($\textcolor{blue}{\triangledown}$1.7, $\textcolor{blue}{\blacktriangledown}$0.0) & $\textcolor{blue}{\triangledown}$3.7 (1.7) & $\textcolor{blue}{\triangledown}$1.7 ($\textcolor{blue}{\triangledown}$1.7, $\textcolor{blue}{\blacktriangledown}$0.0)\\  
\emph{\restscs} & \cellcolor{gray!20}BB & \cellcolor{gray!20}15.7 & \cellcolor{gray!20}56.7 & \cellcolor{gray!20}0.0 (0.0) & \cellcolor{gray!20}0.0 (0.0, 0.0) & \cellcolor{gray!20}0.0 (0.0) & \cellcolor{gray!20}0.0 (0.0, 0.0)\\  
 & WB & 76.7 & 71.6 & $\textcolor{red}{\blacktriangle}$0.1 (0.4) & $\textcolor{red}{\blacktriangle}$0.1 ($\textcolor{red}{\blacktriangle}$0.1, 0.0) & $\textcolor{red}{\blacktriangle}$0.1 (0.4) & $\textcolor{red}{\blacktriangle}$0.1 ($\textcolor{red}{\blacktriangle}$0.1, 0.0)\\  
\emph{\restcountries} & \cellcolor{gray!20}BB & \cellcolor{gray!20}62.4 & \cellcolor{gray!20}76.0 & \cellcolor{gray!20}0.0 (0.0) & \cellcolor{gray!20}0.0 (0.0, 0.0) & \cellcolor{gray!20}0.0 (0.0) & \cellcolor{gray!20}0.0 (0.0, 0.0)\\  
 & WB & 307.9 & 70.5 & $\textcolor{red}{\blacktriangle}$18.7 (5.2) & $\textcolor{red}{\blacktriangle}$58.0 ($\textcolor{red}{\blacktriangle}$58.0, 0.0) & $\textcolor{red}{\blacktriangle}$18.7 (5.2) & $\textcolor{red}{\blacktriangle}$58.0 ($\textcolor{red}{\blacktriangle}$58.0, 0.0)\\  
\emph{\scoutapi} & \cellcolor{gray!20}BB & \cellcolor{gray!20}267.4 & \cellcolor{gray!20}30.5 & \cellcolor{gray!20}32.7 (5.5) & \cellcolor{gray!20}86.8 (78.9, 7.9) & \cellcolor{gray!20}32.7 (5.5) & \cellcolor{gray!20}86.8 (78.9, 7.9)\\  
 & WB & 145.8 & 35.1 & $\textcolor{blue}{\triangledown}$2.3 (1.7) & $\textcolor{blue}{\triangledown}$3.4 ($\textcolor{blue}{\triangledown}$1.8, $\textcolor{blue}{\triangledown}$1.6) & $\textcolor{blue}{\triangledown}$2.3 (1.7) & $\textcolor{blue}{\triangledown}$3.4 ($\textcolor{blue}{\triangledown}$1.8, $\textcolor{blue}{\triangledown}$1.6)\\  
\emph{\sessionservice} & \cellcolor{gray!20}BB & \cellcolor{gray!20}66.2 & \cellcolor{gray!20}56.7 & \cellcolor{gray!20}0.0 (0.0) & \cellcolor{gray!20}0.0 (0.0, 0.0) & \cellcolor{gray!20}0.0 (0.0) & \cellcolor{gray!20}0.0 (0.0, 0.0)\\  
 & WB & 138.5 & 71.5 & $\textcolor{red}{\blacktriangle}$0.1 (0.3) & $\textcolor{red}{\blacktriangle}$0.2 ($\textcolor{red}{\blacktriangle}$0.2, 0.0) & $\textcolor{red}{\blacktriangle}$0.1 (0.3) & $\textcolor{red}{\blacktriangle}$0.2 ($\textcolor{red}{\blacktriangle}$0.2, 0.0)\\  
\emph{\springactuatordemo} & \cellcolor{gray!20}BB & \cellcolor{gray!20}8.2 & \cellcolor{gray!20}87.1 & \cellcolor{gray!20}0.0 (0.0) & \cellcolor{gray!20}0.0 (0.0, 0.0) & \cellcolor{gray!20}0.0 (0.0) & \cellcolor{gray!20}0.0 (0.0, 0.0)\\  
 & WB & 7.7 & 83.9 & 0.0 (0.0) & 0.0 (0.0, 0.0) & 0.0 (0.0) & 0.0 (0.0, 0.0)\\  
\emph{\springbatchrest} & \cellcolor{gray!20}BB & \cellcolor{gray!20}12.1 & \cellcolor{gray!20}34.2 & \cellcolor{gray!20}0.0 (0.0) & \cellcolor{gray!20}0.0 (0.0, 0.0) & \cellcolor{gray!20}0.0 (0.0) & \cellcolor{gray!20}0.0 (0.0, 0.0)\\  
 & WB & 22.2 & 66.2 & $\textcolor{red}{\blacktriangle}$31.1 (5.6) & $\textcolor{red}{\blacktriangle}$6.9 ($\textcolor{red}{\blacktriangle}$6.9, 0.0) & $\textcolor{red}{\blacktriangle}$31.1 (5.6) & $\textcolor{red}{\blacktriangle}$6.9 ($\textcolor{red}{\blacktriangle}$6.8, $\textcolor{red}{\blacktriangle}$0.1)\\  
\emph{\springecommerce} & \cellcolor{gray!20}BB & \cellcolor{gray!20}62.1 & \cellcolor{gray!20}28.0 & \cellcolor{gray!20}2.1 (0.8) & \cellcolor{gray!20}1.3 (1.0, 0.3) & \cellcolor{gray!20}1.6 (0.0) & \cellcolor{gray!20}1.0 (1.0, 0.0)\\  
 & WB & 121.7 & 42.8 & $\textcolor{red}{\triangle}$3.9 (1.5) & $\textcolor{red}{\triangle}$4.8 ($\textcolor{red}{\triangle}$4.8, $\textcolor{blue}{\blacktriangledown}$0.0) & $\textcolor{red}{\triangle}$5.1 (2.0) & $\textcolor{red}{\triangle}$6.3 ($\textcolor{red}{\triangle}$6.3, 0.0)\\  
\emph{\springrestexample} & \cellcolor{gray!20}BB & \cellcolor{gray!20}19.5 & \cellcolor{gray!20}48.1 & \cellcolor{gray!20}0.5 (1.6) & \cellcolor{gray!20}0.1 (0.1, 0.0) & \cellcolor{gray!20}0.5 (1.6) & \cellcolor{gray!20}0.1 (0.1, 0.0)\\  
 & WB & 128.5 & 58.4 & $\textcolor{red}{\triangle}$3.5 (4.6) & $\textcolor{red}{\triangle}$4.6 ($\textcolor{red}{\triangle}$4.6, 0.0) & $\textcolor{red}{\triangle}$3.5 (4.6) & $\textcolor{red}{\triangle}$4.6 ($\textcolor{red}{\triangle}$4.6, 0.0)\\  
\emph{\swaggerpetstore} & \cellcolor{gray!20}BB & \cellcolor{gray!20}53.1 & \cellcolor{gray!20}66.6 & \cellcolor{gray!20}39.0 (2.8) & \cellcolor{gray!20}20.7 (18.4, 2.3) & \cellcolor{gray!20}39.0 (2.8) & \cellcolor{gray!20}20.7 (18.4, 2.3)\\  
 & WB & 84.7 & 68.3 & $\textcolor{blue}{\triangledown}$17.4 (4.3) & $\textcolor{blue}{\triangledown}$14.7 ($\textcolor{blue}{\triangledown}$14.7, $\textcolor{blue}{\blacktriangledown}$0.0) & $\textcolor{blue}{\triangledown}$17.4 (4.3) & $\textcolor{blue}{\triangledown}$14.7 ($\textcolor{blue}{\triangledown}$14.7, $\textcolor{blue}{\blacktriangledown}$0.0)\\  
\emph{\tiltaksgjennomforing} & \cellcolor{gray!20}BB & \cellcolor{gray!20}429.9 & \cellcolor{gray!20}9.0 & \cellcolor{gray!20}0.0 (0.0) & \cellcolor{gray!20}0.0 (0.0, 0.0) & \cellcolor{gray!20}0.0 (0.0) & \cellcolor{gray!20}0.0 (0.0, 0.0)\\  
 & WB & 84.3 & 9.6 & 0.0 (0.0) & 0.0 (0.0, 0.0) & 0.0 (0.0) & 0.0 (0.0, 0.0)\\  
\emph{\trackingsystem} & \cellcolor{gray!20}BB & \cellcolor{gray!20}209.8 & \cellcolor{gray!20}33.9 & \cellcolor{gray!20}74.4 (1.0) & \cellcolor{gray!20}156.2 (147.6, 8.6) & \cellcolor{gray!20}74.4 (1.0) & \cellcolor{gray!20}156.2 (147.7, 8.5)\\  
 & WB & 201.4 & 35.4 & $\textcolor{blue}{\triangledown}$46.8 (13.6) & $\textcolor{blue}{\triangledown}$101.9 ($\textcolor{blue}{\triangledown}$94.7, $\textcolor{blue}{\triangledown}$7.2) & $\textcolor{blue}{\triangledown}$46.8 (13.6) & $\textcolor{blue}{\triangledown}$101.9 ($\textcolor{blue}{\triangledown}$94.7, $\textcolor{blue}{\triangledown}$7.2)\\  
\emph{\usermanagement} & \cellcolor{gray!20}BB & \cellcolor{gray!20}45.9 & \cellcolor{gray!20}38.1 & \cellcolor{gray!20}38.7 (3.0) & \cellcolor{gray!20}17.8 (17.6, 0.2) & \cellcolor{gray!20}38.7 (3.0) & \cellcolor{gray!20}17.8 (17.6, 0.2)\\  
 & WB & 67.5 & 45.1 & $\textcolor{blue}{\triangledown}$8.9 (4.8) & $\textcolor{blue}{\triangledown}$6.1 ($\textcolor{blue}{\triangledown}$6.1, $\textcolor{blue}{\blacktriangledown}$0.0) & $\textcolor{blue}{\triangledown}$8.9 (4.8) & $\textcolor{blue}{\triangledown}$6.1 ($\textcolor{blue}{\triangledown}$6.1, $\textcolor{blue}{\blacktriangledown}$0.0)\\  
\emph{\webgoat} & \cellcolor{gray!20}BB & \cellcolor{gray!20}619.2 & \cellcolor{gray!20}52.0 & \cellcolor{gray!20}36.6 (0.2) & \cellcolor{gray!20}226.5 (224.6, 1.9) & \cellcolor{gray!20}37.9 (0.3) & \cellcolor{gray!20}234.6 (232.7, 1.9)\\  
 & WB & 262.1 & 22.2 & $\textcolor{blue}{\triangledown}$0.4 (0.0) & $\textcolor{blue}{\triangledown}$1.0 ($\textcolor{blue}{\triangledown}$1.0, $\textcolor{blue}{\blacktriangledown}$0.0) & $\textcolor{blue}{\triangledown}$0.4 (0.0) & $\textcolor{blue}{\triangledown}$1.0 ($\textcolor{blue}{\triangledown}$1.0, $\textcolor{blue}{\blacktriangledown}$0.0)\\  
\emph{\youtubemock} & \cellcolor{gray!20}BB & \cellcolor{gray!20}21.2 & \cellcolor{gray!20}45.6 & \cellcolor{gray!20}0.9 (2.9) & \cellcolor{gray!20}0.1 (0.1, 0.0) & \cellcolor{gray!20}0.9 (2.9) & \cellcolor{gray!20}0.1 (0.1, 0.0)\\  
 & WB & 24.5 & 43.7 & $\textcolor{blue}{\blacktriangledown}$0.0 (0.0) & $\textcolor{blue}{\blacktriangledown}$0.0 ($\textcolor{blue}{\blacktriangledown}$0.0, 0.0) & $\textcolor{blue}{\blacktriangledown}$0.0 (0.0) & $\textcolor{blue}{\blacktriangledown}$0.0 ($\textcolor{blue}{\blacktriangledown}$0.0, 0.0)\\  
\midrule 
\#\emph{API} & \cellcolor{gray!20}$\textcolor{red}{\triangle}$ & \cellcolor{gray!20} & \cellcolor{gray!20} & \cellcolor{gray!20} & \cellcolor{gray!20}11 (10, 10) & \cellcolor{gray!20} & \cellcolor{gray!20}16 (17, 12)\\  
 & $\textcolor{blue}{\triangledown}$ &  &  &  & 16 (15, 15) &  & 13 (11, 14)\\  
 & \cellcolor{gray!20}BB & \cellcolor{gray!20} & \cellcolor{gray!20} & \cellcolor{gray!20} & \cellcolor{gray!20}25 (24, 17) & \cellcolor{gray!20} & \cellcolor{gray!20}24 (24, 16)\\  
 & WB &  &  &  & 30 (29, 9) &  & 30 (30, 9)\\  
 & \cellcolor{gray!20}ALL & \cellcolor{gray!20} & \cellcolor{gray!20} & \cellcolor{gray!20} & \cellcolor{gray!20}31 (30, 20) & \cellcolor{gray!20} & \cellcolor{gray!20}31 (31, 20)\\  
\bottomrule 
\end{tabular} 

	}
\end{table}

By comparing two execution environments, flakiness results are similar for most APIs in both modes. 
However, several APIs exhibit noticeable differences, suggesting that flakiness can be strongly influenced by the environment. 
More specifically, \execenv yields higher failure rates than \fuzzenv for WB on 
\blogapi, 
\catwatch, 
\familiebasak, 
\httppatchspring, 
\languagetool, 
\market, 
\ocvn, 
\proxyprint, and
\reservationsapi, as well as for BB on \familiebasak, \microcks, and \webgoat. 
The largest differences are observed for \proxyprint in WB, whose failure rate increases from 2.9\% in \fuzzenv to 45.4\% in \execenv, 
\reservationsapi in WB, which rises from 37.9\% to 63.8\%,
and \ocvn in WB, which increases from 55.2\% to 67.3\%.
In contrast, cases where \fuzzenv yields higher failure rates than \execenv are less common. 
These mainly occur for \genomenexus in both BB and WB, and \paypublicapi in BB. 
Among them, the most substantial differences are observed for \paypublicapi in BB, whose failure rate drops from 66.7\% in \fuzzenv to 0.0\% in \execenv.

When comparing white-box and black-box tests, we observed flakiness in more APIs under WB than under BB, i.e., 30 vs. 26 on \fuzzenv, 30 vs. 25 on \execenv.
This may indicate that WB can expose additional flaky behaviors that remain hidden to BB.
For instance, there are six APIs, i.e.,  \languagetool, \ohsomeapi, \restscs, \restncs, \sessionservice, and \springbatchrest, where BB shows no flakiness but WB introduces non-deterministic behavior.
In addition, we also observed higher failure rate on \ocvn (i.e., rises from 28.6\% in BB to 55.2\% in \fuzzenv and 67.2\% in \execenv), \httppatchspring (from 25.4\% to 38.1\% and 46.1\%), and \reservationsapi (12.6\% to 37.9\% in \fuzzenv and 23.1\% in \execenv). 
On these APIs, except for \ohsomeapi, WB achieved higher code coverage, suggesting that it exercises deeper behaviors that are more likely to trigger non-deterministic failures.
However, among flaky APIs, WB more often reduces flakiness in \fuzzenv, but its effect is more balanced in \execenv.
Specifically, in \fuzzenv, WB decreases the failure rate for 16 APIs and increases it for 11.
In \execenv, WB decreases it for 13 APIs and increases it for 16.
Several APIs show substantially lower flakiness under WB than under BB. 
Representative examples include \restnews (40.3\% vs. 3.7\%), \scoutapi (32.7\% vs. 2.3\%), \microcks (13.7\%/14.8\% vs.~1.5\%/2.5\%), \usermanagement (38.7\% vs. 8.9\%), \webgoat (36.6/37.9\% vs. 0.4\%), and \proxyprint in \fuzzenv (83.8\% vs. 2.9\%). 
There also exist a small number of flaky cases appear only in BB, i.e., on 
\youtubemock, BB exhibits non-zero failure rates while WB eliminates flakiness entirely.

By looking at the results of stability of flakiness, we found that WB exhibits fewer unstable flaky cases than BB, i.e., 9 vs.~17 in \fuzzenv, and 9 vs.~16 in \execenv.
One plausible explanation is that white-box tests can better exercise internal states and hidden dependencies, leading to failures that occur more consistently during test execution.

\begin{results}[Findings of RQ1]
	Flaky tests are widespread in REST APIs under both WB and BB modes. 
	Across 36 APIs, flakiness was observed in 31 APIs in both execution environments, showing that it is common rather than exceptional. 
	WB exhibits flaky behavior more frequently than BB, affecting 30 APIs \textit{vs.} 25 in \fuzzenv and 30 \textit{vs.} 24 in \execenv, which suggests that WB can expose non-determinism hidden from BB. 
	However, this does not mean WB always increases flakiness: among already flaky APIs, WB reduces failure rates for many cases. Moreover, WB yields nearly half as many unstable flaky cases as BB under both runtime environments, indicating that WB more often exposes failures consistently, whereas BB more frequently produces intermittently failing tests.
\end{results}

\subsection{Flakiness Analysis }
\label{subsec:analysis}

We analyzed failing tests by manually inspecting the source code of these test cases and debugging them (i.e., running them in debug-mode inside an IDE)  against the corresponding APIs.
Based on the results of RQ1, we observed flakiness in 31 of the 36 APIs.
As WB exposes more flaky cases and produces more stable tests, which are easier to debug, we selected, for each API, the WB run with the largest number of flaky tests for detailed analysis.
For the API where flakiness was eliminated by \evo in WB mode, we instead analyzed the BB tests.
In total, we performed manual analysis on 2991 failed WB tests (i.e., 1297 on \fuzzenv, and 1694 on \execenv), and two failed BB tests (i.e., 1 on \fuzzenv, and 1 on \execenv). 
Based on the analysis, we identified 9 categories of flakiness sources across the 31 studied APIs, as summarized in Table~\ref{tab:flakiness-types}.

\begin{table}
	\small
	\centering
	\caption{Classification of test flakiness sources}
	\label{tab:flakiness-types}
	\resizebox{.99\linewidth}{!}{
		
\begin{tabular}{p{.4\linewidth} p{.8\linewidth}}
    \toprule 
	\textbf{Category} & \textbf{Description} \\
    \midrule 
	Time-Dependent (\Time) &
	Business logic or fields in responses that depend on the current system time (e.g., timestamps). \\
	
	Randomness-Dependent (\Rand) &
	Nondeterministic values such as UUIDs and random seeds. \\
	
	Cryptographic Validation (\Crypt) &
	Domain-specific verification logic (e.g., hashing and encryption), primarily semantic and partially influenced by randomness. \\
	
	Unordered Collection (\Unord) &
	Assertions assume a fixed element order in collections with undefined iteration order (e.g., \texttt{Set}, \texttt{HashSet}), or undocumented ordering guarantees. \\
	
	Runtime-Dependent Message (\RunMsg) &
	Dynamically generated error messages containing runtime-dependent elements, such as memory addresses, stack traces, instance identifiers, reflection-related information, and \texttt{InputStream} references. \\
	
	Stateful Resource (\DynState) &
	Dependence on mutable persistent or in-memory state, e.g., databases, caches, in-memory stores, and static singletons. \\
	
	Runtime Environment (\Env) &
	Variations in execution environment, including locale and language settings, JAR metadata, classpath scanning, module loading, and framework-managed endpoints, and runtime configuration values derived from the host environment, such as cache locations, temporary directories, and filesystem paths.\\
	
	Unclassified (\Unk) &
	Failures whose root causes could not be clearly identified based on current investigation. \\

	Generation Error  (\GenErr) &
	Failures caused by improperly generated test cases, such as incorrect request paths, invalid request sequences, unsatisfied state dependencies, brittle assertions, environment-dependent interactions, or insufficient cleanup between tests. \\
	
    \bottomrule 
\end{tabular}

	}
	
\end{table}




\subsubsection{\Env}
Overall, \Env is the most dominant category in terms of the number of flaky tests, accounting for 1,161 flaky tests in 15 APIs.
This dominance is mainly driven by a few APIs with a very large number of environment-related flaky tests, most notably \ocvn (906) and \proxyprint (175).
However, we found that, on these two APIs, all 1081 flaky tests occur because the test asserts against a localized validation message whose exact text depends on the runtime environment \texttt{Locale}. 
The same \texttt{@Pattern} constraint can produce different default messages across environments. 
In an English locale, the message appears as \texttt{"must match "\^{}[a-zA-Z0-9]*\$""}, whereas under a different locale, the phrase \texttt{"must match"} is replaced by its localized equivalent.
As a result, the test may pass when executed under one locale and fail under another, even though the validation logic itself behaves identically.
On \languagetool and \catwatch, we also observed sources of flakiness that depend on properties of the execution environment, such as classpath resource descriptors and temporary folders as examples:
\begin{lstlisting}[numbers=none,language=Kotlin,basicstyle=\scriptsize\ttfamily,escapeinside={(*@}{@*)}]
// (*@\languagetool@*)languagetool
.post(baseUrlOfSut + "/v2/check")
.then()
.statusCode(400)
.assertThat()
.body(containsString("Error: 'KPj5J' is not a language code known to LanguageTool. Supported language codes are: ar, ast-ES, be-BY, br-FR, ca-ES, ca-ES-valencia, da-DK, de, de-AT, de-CH, de-DE, de-DE-x-simple-language, el-GR, en, en-AU, en-CA, en-GB, en-NZ, en-US, en-ZA, eo, es, fa, fr, ga-IE, gl-ES, it, ja-JP, km-KH, pl-PL, pt, pt-AO, pt-BR, pt-MZ, pt-PT, ro-RO, ru-RU, sk-SK, sl-SI, sv, ta-IN, tl-PH, uk-UA, zh-CN. The list of languages is read from META-INF/org/languagetool/language-module.properties in the Java classpath. See https://dev.languagetool.org/java-api for details."));
// (*@\catwatch@*)catwatch
.get(baseUrlOfSut + "/config?EMextraParam123=42")
.then()
.statusCode(200)
.assertThat()
.contentType("application/json")
.body("'cache.path'", containsString("/home/user/workspace/temp/tmp_catwatch/cache_10062"))
\end{lstlisting}
Moreover, we observed additional flakiness in framework-provided endpoints such as \texttt{/actuator/health} on \springecommerce. 
The structure and semantics of these responses often depend on runtime configuration, registered health contributors, security settings, and the current state of external infrastructure, which can lead to flaky tests when the execution environment differs.

\subsubsection{\RunMsg}
\RunMsg is the second largest source in terms of affected APIs, i.e., across 11 APIs. 
One main cause of \RunMsg relates to runtime-dependent elements in response messages as illustrated in the example below:
\begin{lstlisting}[numbers=none,language=Kotlin,basicstyle=\scriptsize\ttfamily,escapeinside={(*@}{@*)}]
// (*@\proxyprint@*) proxyprint
.body(" { " + 
	" \"id\": 530, " + 
	" \"roles\": { " + 
		" \"EM_tainted_map\": \"_EM_0_XYZ_\" " + 
		" } " + 
	" } ")
.post(baseUrlOfSut + "/admin/register")
.then()
.statusCode(400)
.assertThat()
.contentType("application/json")
.body("'status'", numberMatches(400.0))
.body("'error'", containsString("Bad Request"))
.body("'exception'", containsString("org.springframework.http.converter.HttpMessageNotReadableException"))
.body("'message'", containsString("Could not read document: Can not deserialize instance of java.util.HashSet out of START_OBJECT token\n at [Source: java.io.ByteArrayInputStream@72c11c70; line: 1, column: 20] (through reference chain: io.github.proxyprint.kitchen.models.Admin[\"roles\"]); nested exception is com.fasterxml.jackson.databind.JsonMappingException: Can not deserialize instance of java.util.HashSet out of START_OBJECT token\n at [Source: java.io.ByteArrayInputStream@72c11c70; line: 1, column: 20] (through reference chain: io.github.proxyprint.kitchen.models.Admin[\"roles\"])"))
.body("'path'", containsString("/admin/register"));
/*
Actual: Could not read document: Can not deserialize instance of java.util.HashSet out of START_OBJECT token\n at [Source: java.io.PushbackInputStream@67d4bd48; line: 1, column: 26] (through reference chain: io.github.proxyprint.kitchen.models.Admin["roles"]); nested exception is com.fasterxml.jackson.databind.JsonMappingException: Can not deserialize instance of java.util.HashSet out of START_OBJECT token\n at [Source: java.io.PushbackInputStream@67d4bd48; line: 1, column: 26] (through reference chain: io.github.proxyprint.kitchen.models.Admin["roles"])
*/
\end{lstlisting}
This example also exhibits flakiness relating to the reported exception source.
During test generation, the JSON is likely deserialized directly from an in-memory byte array, which leads to \texttt{ByteArrayInputStream} appearing in the exception message. 
During actual execution, by contrast, the framework reads the payload from the HTTP request body, where it may be wrapped in additional stream layers such as \texttt{PushbackInputStream}.
As a result, the same deserialization error can produce slightly different exception texts across runs.
This source of flakiness affects various APIs (i.e., \gestaohospital, \market, \trackingsystem, \restnews, \proxyprint, and \usermanagement), as JSON parsing is a fundamental mechanism used throughout REST APIs.

\subsubsection{\DynState}
\DynState is also common in the open-source APIs, affecting 11 APIs and accounting for 130 flaky tests. 
We found that, on 7 APIs, their flaky assertions are primarily due to internal state stored in databases. 
White-box fuzzers may mitigate this source of flakiness if they are able to handle database state effectively. 
However, this depends on the capability of the specific fuzzer. 
In our study, for example, \evo failed to mitigate database-related flakiness in 7 APIs (i.e., \featuresservice, \springbatchrest, \market, \springrestexample, \swaggerpetstore, \trackingsystem and \usermanagement), although it was effective for the remaining 18 database-interacting APIs.
For black-box fuzzers, by contrast, such cases are generally not feasible to handle.
Another typical source of \DynState flakiness is external dependencies, but this is handled by WB \evo and was therefore not observed in our analysis. 
Moreover, we identified two APIs, namely \httppatchspring and \quartzmanager, whose flaky behavior is related to in-memory state, as examples illustrated below:
\begin{lstlisting}[numbers=none,language=java,basicstyle=\scriptsize\ttfamily,escapeinside={(*@}{@*)}]
// (*@\httppatchspring@*) http-patch-spring
private static final List<Contact> CONTACTS = new ArrayList<>();

// (*@\quartzmanager@*) quartz-manager
@Getter
private List<Class<? extends AbstractQuartzManagerJob>> jobClasses = new ArrayList<>();
private List<String> jobClassPackages = new ArrayList<>();
\end{lstlisting}

\begin{table}
	\small
	\caption{The number of flaky tests for each API across the nine flakiness categories on \fuzzenv (\textit{F}) and \execenv (\textit{E}). }
	\label{tab:rq1_types}
	\squeezeupfigureAbove
	\resizebox{.99\linewidth}{!}{
		\begin{tabular}{l rrrr rrrrr r}\\ 
\toprule 
SUT &  \Time & \Rand & \Crypt & \Unord & \RunMsg & \DynState & \Env & \Unk & \GenErr & Total  \\ 
 &  (\emph{F}, \emph{E}) & (\emph{F}, \emph{E}) & (\emph{F}, \emph{E}) & (\emph{F}, \emph{E}) & (\emph{F}, \emph{E}) & (\emph{F}, \emph{E}) & (\emph{F}, \emph{E}) & (\emph{F}, \emph{E}) & (\emph{F}, \emph{E}) &   \\ 
\midrule 
\emph{\blogapi} & (10, 6) & (0, 0) & (0, 0) & (0, 0) & (0, 0) & (0, 0) & (0, 21) & (0, 0) & (0, 0) & (1, 2) \\ 
\emph{\catwatch} & (0, 0) & (0, 0) & (0, 0) & (0, 0) & (0, 0) & (0, 0) & (11, 11) & (0, 56) & (0, 0) & (1, 2) \\ 
\emph{\cwaverification} & (0, 0) & (0, 0) & (0, 0) & (0, 0) & (0, 0) & (0, 0) & (0, 0) & (0, 0) & (5, 5) & (1, 1) \\ 
\emph{\familiebasak} & (0, 0) & (0, 0) & (0, 0) & (0, 0) & (0, 1) & (0, 0) & (2, 11) & (0, 0) & (0, 0) & (1, 2) \\ 
\emph{\featuresservice} & (0, 0) & (0, 0) & (0, 0) & (0, 0) & (0, 0) & (1, 1) & (0, 0) & (0, 0) & (0, 0) & (1, 1) \\ 
\emph{\genomenexus} & (0, 0) & (0, 0) & (0, 0) & (0, 0) & (0, 0) & (0, 0) & (3, 3) & (23, 23) & (0, 0) & (2, 2) \\ 
\emph{\gestaohospital} & (0, 0) & (0, 0) & (0, 0) & (0, 0) & (16, 16) & (0, 0) & (2, 2) & (0, 0) & (0, 0) & (2, 2) \\ 
\emph{\httppatchspring} & (0, 0) & (0, 0) & (0, 0) & (0, 0) & (0, 0) & (13, 13) & (0, 2) & (0, 0) & (0, 0) & (1, 2) \\ 
\emph{\languagetool} & (0, 0) & (0, 0) & (0, 0) & (0, 0) & (0, 0) & (0, 0) & (1, 3) & (0, 0) & (0, 0) & (1, 1) \\ 
\emph{\market} & (0, 0) & (0, 0) & (0, 0) & (4, 4) & (7, 7) & (1, 1) & (0, 7) & (0, 0) & (0, 0) & (3, 4) \\ 
\emph{\microcks} & (2, 2) & (0, 0) & (0, 0) & (0, 0) & (0, 0) & (0, 0) & (1, 1) & (0, 0) & (0, 0) & (2, 2) \\ 
\emph{\ocvn} & (0, 0) & (0, 0) & (0, 0) & (753, 906) & (0, 0) & (0, 0) & (0, 906) & (0, 0) & (0, 0) & (1, 2) \\ 
\emph{\ohsomeapi} & (0, 0) & (0, 0) & (0, 0) & (0, 0) & (2, 2) & (0, 0) & (0, 0) & (0, 0) & (3, 3) & (2, 2) \\ 
\emph{\paypublicapi} & (0, 0) & (0, 0) & (0, 0) & (0, 0) & (0, 0) & (0, 0) & (2, 2) & (0, 0) & (0, 0) & (1, 1) \\ 
\emph{\personcontroller} & (0, 0) & (0, 0) & (0, 0) & (0, 0) & (0, 0) & (0, 0) & (11, 11) & (0, 0) & (0, 0) & (1, 1) \\ 
\emph{\proxyprint} & (18, 1) & (0, 0) & (0, 0) & (0, 0) & (7, 5) & (0, 3) & (0, 175) & (0, 0) & (0, 0) & (2, 4) \\ 
\emph{\quartzmanager} & (0, 0) & (0, 0) & (0, 0) & (0, 0) & (0, 0) & (2, 2) & (0, 0) & (13, 13) & (0, 0) & (2, 2) \\ 
\emph{\reservationsapi} & (0, 0) & (0, 0) & (0, 0) & (0, 0) & (5, 5) & (0, 0) & (0, 0) & (0, 0) & (0, 0) & (1, 1) \\ 
\emph{\restnews} & (0, 0) & (0, 0) & (0, 0) & (0, 0) & (4, 4) & (0, 0) & (0, 0) & (0, 0) & (0, 0) & (1, 1) \\ 
\emph{\restscs} & (0, 0) & (0, 0) & (0, 0) & (0, 0) & (0, 0) & (0, 0) & (0, 0) & (0, 0) & (1, 1) & (1, 1) \\ 
\emph{\restcountries} & (0, 0) & (0, 0) & (0, 0) & (0, 0) & (0, 0) & (0, 0) & (0, 0) & (0, 0) & (89, 89) & (1, 1) \\ 
\emph{\scoutapi} & (5, 5) & (0, 0) & (0, 0) & (3, 3) & (0, 0) & (0, 0) & (0, 0) & (0, 0) & (1, 1) & (3, 3) \\ 
\emph{\sessionservice} & (0, 0) & (1, 1) & (0, 0) & (0, 0) & (0, 0) & (0, 0) & (0, 0) & (0, 0) & (0, 0) & (1, 1) \\ 
\emph{\springbatchrest} & (0, 0) & (0, 0) & (0, 0) & (0, 0) & (0, 0) & (9, 9) & (0, 0) & (0, 0) & (0, 0) & (1, 1) \\ 
\emph{\springecommerce} & (0, 0) & (0, 0) & (0, 0) & (0, 0) & (0, 0) & (0, 0) & (7, 10) & (0, 0) & (0, 0) & (1, 1) \\ 
\emph{\springrestexample} & (0, 0) & (0, 0) & (0, 0) & (0, 0) & (0, 0) & (16, 16) & (0, 0) & (0, 0) & (0, 0) & (1, 1) \\ 
\emph{\swaggerpetstore} & (3, 3) & (0, 0) & (0, 0) & (0, 0) & (2, 2) & (7, 7) & (0, 0) & (8, 8) & (0, 0) & (4, 4) \\ 
\emph{\trackingsystem} & (0, 0) & (0, 0) & (1, 0) & (12, 11) & (34, 46) & (69, 68) & (0, 0) & (34, 35) & (0, 0) & (5, 4) \\ 
\emph{\usermanagement} & (0, 0) & (1, 1) & (0, 0) & (0, 0) & (4, 4) & (10, 10) & (0, 0) & (4, 4) & (0, 0) & (4, 4) \\ 
\emph{\webgoat} & (0, 0) & (0, 0) & (0, 0) & (0, 0) & (0, 0) & (0, 0) & (1, 1) & (0, 0) & (0, 0) & (1, 1) \\ 
\emph{\youtubemock} & (0, 0) & (1, 1) & (0, 0) & (0, 0) & (0, 0) & (0, 0) & (0, 0) & (0, 0) & (0, 0) & (1, 1) \\ 
\midrule 
\#F & (38, 17) & (3, 3) & (1, 0) & (772, 924) & (81, 92) & (128, 130) & (41, 1166) & (82, 139) & (99, 99) & \\ 
\#APIs & (5, 5) & (3, 3) & (1, 0) & (4, 4) & (9, 10) & (9, 10) & (10, 15) & (5, 6) & (5, 5) & \\ 
\bottomrule 
\end{tabular} 

	}
\end{table}

\subsubsection{\Unord}
We observed it in 4 APIs, yet it accounts for 924 flaky tests.
The main source of this flakiness is unstable ordering in the response content, which may be caused either by nondeterministic ordering in the underlying data source or by unordered collections returned by the API, as illustrated by the examples below:
\begin{lstlisting}[numbers=none,language=java,basicstyle=\scriptsize\ttfamily,escapeinside={(*@}{@*)}]
// (*@\scoutapi@*) scount-api
.get(baseUrlOfSut + "/api/v1/users?EMextraParam123=42")
.then()
.statusCode(200)
.assertThat()
.contentType("application/json")
.body("size()", equalTo(3))
.body("[0].'name'", containsString("INTEGRATION TEST MODERATOR"))

// (*@\ocvn@*) ocvn
.body("'exception'", containsString("org.springframework.validation.BindException"))
...
.body("'errors'[0].'codes'", hasItems("EachPattern.yearFilterPagingRequest.contrMethod[1]", "EachPattern.yearFilterPagingRequest.contrMethod", "EachPattern.contrMethod[1]", "EachPattern.contrMethod", "EachPattern.java.lang.String", "EachPattern"))
\end{lstlisting}
Upon further investigation, we found that 914 of these flaky tests on \market, \ocvn and \trackingsystem (as the above example in \ocvn) were caused by unordered error messages in the responses.

\subsubsection{\Time, \Rand,  \Crypt and \Unk}
Several categories are less frequent but still recurrent across APIs.
\Time appears in 5 APIs with 38 flaky tests, and \Rand appears in 3 APIs with 3 flaky tests.
These categories are thus relatively common in terms of API coverage, but they typically contribute to only a limited number of flaky tests per API.
For example, \Time-related flakiness is spread across APIs such as \blogapi, \microcks, \proxyprint, \scoutapi, and \swaggerpetstore, while \Rand-related flakiness is observed in 
\sessionservice, \usermanagement, and \webgoat.
\Crypt is the rarest category that appears in only one API, i.e., \trackingsystem, indicating that it is not a major source of flakiness in open-source REST APIs.
\begin{lstlisting}[numbers=none,language=Kotlin,basicstyle=\scriptsize\ttfamily,escapeinside={(*@}{@*)}]
.body(" { " +
	" \"credentialId\": 908, " +
	" \"enabled\": true, " +
	" \"password\": \"_EM_2055_XYZ_\", " +
	" \"role\": \"OU\", " +
	" \"username\": \"iFKw1\" " +
	" } ")
.put(baseUrlOfSut + "/app/api/credentials?" +
	"password=Nm2Z2&" +
	"username=R")
.then()
.statusCode(200)
.assertThat()
.contentType("application/json")
.body("'credentialId'", numberMatches(16.0))
.body("'username'", containsString("iFKw1"))
.body("'password'", containsString("$2a$10$dpyE.RibfQza7.9TD65vT.dzi.OGm2VzqKDYNjMMkIH7obCbsD6.W"))
.body("'enabled'", equalTo(true))
.body("'role'", containsString("OU"));

/*
java.lang.AssertionError: 1 expectation failed.
JSON path 'password' doesn't match.
Expected: a string containing "$2a$10$dpyE.RibfQza7.9TD65vT.dzi.OGm2VzqKDYNjMMkIH7obCbsD6.W"
Actual: $2a$10$Nv9OKJP1TjI9uQfwWdYZsumNi0tLOC2a/q5Dco4klHcOHsUZVACQi
*/
\end{lstlisting}

In our analysis, we also encountered cases whose sources of flakiness remain unclear (\Unk), as we were unable to reproduce them during debugging.
Nevertheless, this category still appears in 6 APIs and accounts for 139 flaky tests, for example in \catwatch (56).
As illustrated by the example below, it is unclear how the system state evolves during fuzzing and why the same request yields different outcomes across executions.
\begin{lstlisting}[numbers=none,language=Kotlin,basicstyle=\scriptsize\ttfamily,escapeinside={(*@}{@*)}]
// (*@\catwatch@*) catwatch
.get(baseUrlOfSut + "/statistics/projects?" + 
"start_date=dhkj3TsEAFY&" + 
"access_token=&" + 
"offset=Zn5IQ&" + 
"language=gR")
.then()
.statusCode(400)
.assertThat()
.contentType("application/json")
.body("'error'", containsString("invalid_request"))
.body("'error_description'", containsString("Access Token not valid"));
/*
1 expectation failed.
Expected status code <400> but was <500>.
*/

// (*@\quartzmanager@*) quartz-manager
.get(baseUrlOfSut + "/quartz-manager/scheduler/run?EMextraParam123=_EM_21_XYZ_")
.then()
.statusCode(500)// it/fabioformosa/quartzmanager/api/services/SchedulerService_25_start
.assertThat()
.contentType("application/json")
.body("'status'", numberMatches(500.0))
.body("'error'", containsString("Internal Server Error"))
.body("'path'", containsString("/quartz-manager/scheduler/run"));
/*
java.lang.AssertionError: 1 expectation failed.
Expected status code <500> but was <204>.
*/
\end{lstlisting}
This suggests that identifying the root causes of flakiness in REST APIs remains challenging, because failures may depend on system states that are reached during fuzzing but are difficult to reconstruct exactly during later execution.

\subsubsection{\GenErr}
Besides REST API flakiness, the fuzzer itself can also introduce failing tests and false positives that are mistakenly identified as flaky tests.
In this study,  \GenErr appears in 4 APIs, but when it occurs, it can be substantial, as shown by \restcountries with 89 flaky tests.
\begin{lstlisting}[numbers=none,language=Kotlin,basicstyle=\scriptsize\ttfamily,escapeinside={(*@}{@*)}]
.get(baseUrlOfSut + "/rest/v2/capital/the%20valley?fields=4f185dVbVe")
.then()
.statusCode(200)
.assertThat()
.contentType("application/json")
.body("size()", equalTo(1))
.body("'[0]'.isEmpty()", is(true)); // "[0].isEmpty()"
\end{lstlisting}
The assertion fails because \texttt{`[0]'} is treated as a string, not as the first JSON array element. 
So the code checks whether the string \texttt{"[0]"} is empty, which is false.

Among the APIs, 16 are affected by \textit{multiple sources of flakiness} and the other 15 are impacted by a single source.
For example, \trackingsystem involves five categories (\Crypt, \Unord, \RunMsg, \DynState, and \Unk), while 
\market, \proxyprint, \swaggerpetstore, and \usermanagement each involve four categories.
Overall, these results suggest that flakiness in REST APIs is heterogeneous in nature: some APIs are primarily affected by a single source, whereas many others are influenced by multiple interacting sources. 
Among the sources, environment-related issues (\Env) and runtime-dependent outputs (\RunMsg and \Unord) are recurring sources of flakiness and account for a large share of flaky tests in the open-source APIs.
This may also relate to fuzzers that can expose error scenarios, thereby increasing the likelihood of observing flaky behavior.


\begin{results}[Findings of RQ2]
	The primary sources of flakiness in REST APIs are runtime environment variations (\Env), stateful resources (\DynState), unordered responses (\Unord), and runtime-dependent messages (\RunMsg). 
	Among them, \Env is the dominant source in terms of the number of flaky tests, followed by \Unord, while \DynState and \RunMsg are also common across APIs. 
	Other causes, such as \Time, \Rand, appear less frequently, and \Crypt is rare.
	In addition, flakiness in REST APIs is not caused by a single dominant factor, but rather arises from a combination of diverse sources.
\end{results}

\section{Detection and Handling of Flaky Tests}
\label{sec:flakyhandling}

Based on our flakiness analysis results, we propose a post-processing approach for detecting and handling flaky tests, i.e., \method.
With the nine identified sources of flakiness, some, such as \DynState and \Env, generally require access to source code or execution in different environments to diagnose. 
In contrast, others, including \Time, \Rand, \Crypt, and \Unord, can be identified or inferred from observable HTTP responses. 
To be applicable to both black-box and white-box fuzzers, \method relies only on observable HTTP responses and does not require access to source code or internal execution traces. 
This design also makes the approach well suited to industrial settings where implementation visibility is limited.

In the rest of the section, we first discussion the flakiness detection capability of \method (Section~\ref{subsec:detection}) and then its handling strategy of flaky tests in Section~\ref{subsec:handling}. Sections~\ref{subsec:R3Expeirment} and~\ref{subsec:RQ3Results} report the empirical study conducted to evaluate \method and its results.

\begin{figure}[t]
	\includegraphics[width=1\linewidth]{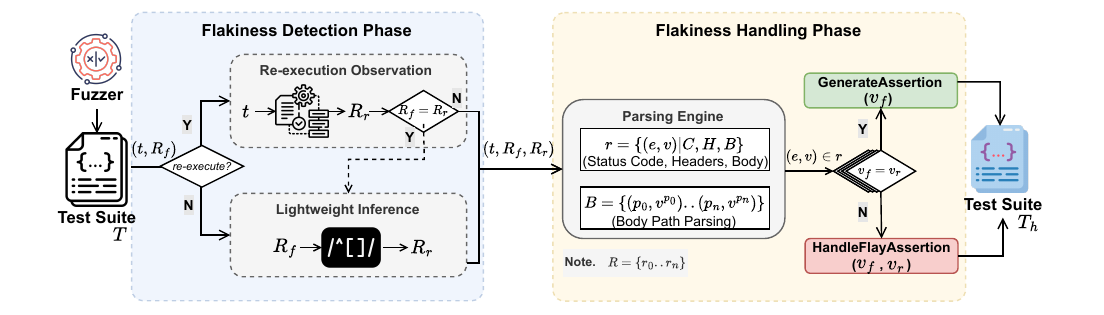}
	\squeezeupfigureAbove
	\caption{The overview of \method}
	\label{fig:method}
	\vspace{-0.3cm}
\end{figure}

\subsection{Flakiness Detection}\label{subsec:detection}

To identify flaky tests, we perform a post-processing phase after fuzzing. All test cases generated by each fuzzer are re-executed under the same configuration and environment. We then compare responses obtained during the fuzzing phase with those collected during the post-processing phase. If the responses differ according to predefined comparison rules, the corresponding test case is marked as potentially flaky.

This approach is based on the assumption that deterministic tests should produce equivalent responses across repeated executions under controlled conditions. Deviations from this behavior indicate potential sources of nondeterminism.
Each HTTP response is represented as a tuple $R = \langle C, H, B \rangle$, where $C$ denotes the status code, $H$ represents the response headers, and $B$ denotes the response body.
Given two responses $R_f$ and $R_r$ obtained during the fuzzing and re-execution phases, respectively, we compare their corresponding elements, including the status code, response headers, and response body.
If any of these comparisons fails, the corresponding flakiness information is recorded and later used for assertion generation.
Figure~\ref{fig:method} presents the response comparison procedure.

\emph{Lightweight Inference.}
As a complement to re-execution-based detection, we introduce a lightweight inference mechanism as a secondary handler. 
This is necessary because re-execution may fail to expose some forms of flakiness. 
For example, random or time-dependent values may remain unchanged if re-execution occurs too close to the original run, and a single repeated execution may still produce the same run-dependent identifier by chance, as we observed in \reservationsapi. 
\vspace{-0.2cm}
\begin{lstlisting}[numbers=none,language=Kotlin,escapeinside={(*@}{@*)}]
"..default message [email],[Ljavax.validation.constraints.Pattern$Flag;@5372cc34,.*];.."
\end{lstlisting}
\vspace{-0.2cm}
Moreover, some unstable response elements may vary only under conditions not exercised by one immediate re-execution. Therefore, when re-execution does not report flakiness, our approach applies a lightweight inference procedure.

The inference mechanism is implemented using pattern-based matching rules that replace potential run-dependent elements with predefined placeholders, e.g., \texttt{\_EM\_POTENTIAL\_OBJECT\_FLAKINESS\_}. 
Currently, our approach targets time-dependent values, randomness-dependent identifiers, cryptographic artifacts, and runtime-dependent messages, following widely adopted standards and specifications.\footnote{ISO 8601 for date and time representations, IEEE POSIX for Unix timestamps, RFC 4122 for UUIDs, RFC 4648 and RFC 7519 for Base64 and JSON Web Tokens, RFC 1321 and FIPS 180-4 for cryptographic hash functions, and the Java Platform Specification for runtime-generated identifiers and stack trace formats}

Note that when re-execution detection is enabled, the inference procedure is only triggered when re-execution does not report flakiness.
However, the inference handler can also be employed independently to reduce the execution overhead associated with repeated test runs.

\subsection{Handling Detected Flaky Tests}\label{subsec:handling}

After identifying potentially flaky test cases, we apply automated post-processing to improve the stability and reusability of the generated test suites. 
Rather than discarding flaky tests entirely, our approach preserves their structural and behavioral coverage while mitigating nondeterministic failures.

Specifically, we further analyze response differences to localize unstable assertions within each test. 
Our goal is to preserve stable checks while disabling only those assertions that depend on nondeterministic response fields.

We handled assertions that validate status code, header and response body content, e.g., \texttt{.body(...)} statements.
In addition, regarding body content,
for each flaky test $t$, we extract all body-related assertions of the form, i.e., \texttt{.body(path, matcher)}, where \texttt{path} denotes a JSON field and \texttt{matcher} specifies the expected value.
During assertion generation, we compare value (i.e., $v_f$ and $v_r$ in 
Figure~\ref{fig:method}) on each path of response bodies obtained in the fuzzing and post-processing phases.
If the value associated with \texttt{path} differs between $v_f$ and $v_r$, the corresponding assertion is commented out and annotated with additional flakiness information, including the path, the value observed during fuzzing, and the value observed during re-execution.
The following illustrates assertions after the post-processing:
\begin{lstlisting}[numbers=none,language=Kotlin,escapeinside={(*@}{@*)}]
(*@\textbf{Original assertion}@*)
.body("'calculatedPastTime'", containsString("2026-12-03T06:38:31.272230"))
(*@\textbf{After post-processing}@*)
// Flaky value of field "'calculatedPastTime'": 2026-12-03T06:38:31.272230 vs. 2026-12-03T06:38:30.713502
// .body("'calculatedPastTime'", containsString("2026-12-03T06:38:31.272230"))
\end{lstlisting}

Figure~\ref{fig:method} presents our procedure for selectively disabling flaky assertions.
As the example, assertions involving volatile fields are replaced with commented statements, while stable assertions are preserved. 
This allows the test to continue validating deterministic aspects of the API behavior.
In addition, all modified assertions are explicitly marked in the generated code to facilitate manual inspection and future refinement by developers. 
This also ensures transparency and prevents the masking of genuine defects.

\subsection{Experiment Settings}\label{subsec:R3Expeirment}

To evaluate the effectiveness of \method for detecting and handling test flakiness, we integrated it into \evo and conducted an experiment to answer the following RQ:

\begin{description}
	\item[{\bf RQ3}:] \rqC
\end{description}

\textbf{Experiment Settings.} 
To study the performance on black-box and white-box tests, 
we enabled our approach in both modes of \evo and ran each mode with a one-hour search budget, repeated 10 times, to generate tests. 
Since flakiness detection was performed only in the fuzzing environment, our approach could not capture environment-dependent flakiness.
Therefore, we executed the generated tests only in \fuzzenv and repeated each test execution 100 times. 

\subsection{Results of Flakiness Detection and Handling}\label{subsec:RQ3Results}

By applying our flakiness-handling strategies, we were able to identify and handle a substantial number of flaky tests across the studied APIs.
In Table~\ref{tab:flakiness_handled}, \#$RF$ denotes the number of resolved flaky assertions in tests, i.e., the assertions identified as flaky and handled by commenting them out.
Note that, to avoid flakiness completely, a trivial solution would be to comment out every single assertion.
A test cannot fail if it does not assert anything.
However, such test cases would become useless for \emph{regression testing} purposes.
Our goal is to comment out only the minimal set of assertions to prevent flakiness, while still trying to maintain test case effectiveness.

\begin{table}
	\small
	\caption{Results for the number of resolved flaky tests (\#$RF$), failure rate ($FR$\%), consistent-failure rate ($FR_c$\%), and unstable-failure rate ($FR_u$\%), together with their comparison results in terms of $\hat{A}_{12}$ and relative improvement.}
	\label{tab:flakiness_handled}
	\squeezeupfigureAbove
	\resizebox{.95\linewidth}{!}{
		\begin{tabular}{ll r | rrr | rrr | rrr}\\ 
\toprule 
SUT & Mode & \#$RF$ & $FR$\% & $\hat{A}_{12}$ & $Rel$\% & $FR_c$\% & $\hat{A}_{12}$ & $Rel$\% & $FR_u$\% & $\hat{A}_{12}$ & $Rel$\%  \\ 
\midrule 
\emph{\blogapi} & \cellcolor{gray!20}BB & \cellcolor{gray!20}0.0 & \cellcolor{gray!20}9.0 & \cellcolor{gray!20}0.5 & \cellcolor{gray!20}0.0 & \cellcolor{gray!20}9.0 & \cellcolor{gray!20}0.5 & \cellcolor{gray!20}0.0 & \cellcolor{gray!20}0.0 & \cellcolor{gray!20}0.5 & \cellcolor{gray!20}NaN \\ 
 & WB & \textbf{14.6} & 0.1 & \textbf{0.0} & -97.1 & 0.1 & \textbf{0.0} & -97.1 & 0.0 & 0.5 & NaN \\ 
\emph{\catwatch} & \cellcolor{gray!20}BB & \cellcolor{gray!20}\textbf{0.3} & \cellcolor{gray!20}5.3 & \cellcolor{gray!20}\textbf{0.2} & \cellcolor{gray!20}-25.6 & \cellcolor{gray!20}2.8 & \cellcolor{gray!20}0.5 & \cellcolor{gray!20}-20.3 & \cellcolor{gray!20}2.5 & \cellcolor{gray!20}0.3 & \cellcolor{gray!20}-30.8 \\ 
 & WB & 0.0 & 4.5 & 0.5 & -5.8 & 4.5 & 0.5 & -5.8 & 0.0 & 0.5 & NaN \\ 
\emph{\cwaverification} & \cellcolor{gray!20}BB & \cellcolor{gray!20}0.0 & \cellcolor{gray!20}100.0 & \cellcolor{gray!20}0.5 & \cellcolor{gray!20}0.0 & \cellcolor{gray!20}100.0 & \cellcolor{gray!20}0.5 & \cellcolor{gray!20}0.0 & \cellcolor{gray!20}0.0 & \cellcolor{gray!20}0.5 & \cellcolor{gray!20}NaN \\ 
 & WB & 0.0 & 100.0 & 0.5 & 0.0 & 100.0 & 0.5 & 0.0 & 0.0 & 0.5 & NaN \\ 
\emph{\familiebasak} & \cellcolor{gray!20}BB & \cellcolor{gray!20}0.0 & \cellcolor{gray!20}0.6 & \cellcolor{gray!20}0.5 & \cellcolor{gray!20}-0.7 & \cellcolor{gray!20}0.6 & \cellcolor{gray!20}0.5 & \cellcolor{gray!20}-0.7 & \cellcolor{gray!20}0.0 & \cellcolor{gray!20}0.5 & \cellcolor{gray!20}NaN \\ 
 & WB & \textbf{7.2} & 1.1 & 0.5 & +0.1 & 1.1 & 0.5 & +0.1 & 0.0 & 0.5 & NaN \\ 
\emph{\featuresservice} & \cellcolor{gray!20}BB & \cellcolor{gray!20}0.0 & \cellcolor{gray!20}6.1 & \cellcolor{gray!20}\textbf{0.9} & \cellcolor{gray!20}+24.4 & \cellcolor{gray!20}3.2 & \cellcolor{gray!20}0.6 & \cellcolor{gray!20}-17.1 & \cellcolor{gray!20}2.9 & \cellcolor{gray!20}0.6 & \cellcolor{gray!20}+171.8 \\ 
 & WB & 0.0 & 0.6 & 0.5 & +5.9 & 0.6 & 0.5 & +5.9 & 0.0 & 0.5 & NaN \\ 
\emph{\genomenexus} & \cellcolor{gray!20}BB & \cellcolor{gray!20}0.0 & \cellcolor{gray!20}29.3 & \cellcolor{gray!20}0.6 & \cellcolor{gray!20}+3.7 & \cellcolor{gray!20}20.6 & \cellcolor{gray!20}\textbf{1.0} & \cellcolor{gray!20}+15.9 & \cellcolor{gray!20}8.7 & \cellcolor{gray!20}0.4 & \cellcolor{gray!20}-17.1 \\ 
 & WB & \textbf{0.1} & 22.0 & 0.4 & -5.1 & 21.0 & 0.4 & -5.7 & 1.0 & 0.5 & +11.0 \\ 
\emph{\gestaohospital} & \cellcolor{gray!20}BB & \cellcolor{gray!20}\textbf{3.6} & \cellcolor{gray!20}27.7 & \cellcolor{gray!20}\textbf{0.1} & \cellcolor{gray!20}-26.1 & \cellcolor{gray!20}23.9 & \cellcolor{gray!20}\textbf{0.1} & \cellcolor{gray!20}-33.9 & \cellcolor{gray!20}3.8 & \cellcolor{gray!20}0.7 & \cellcolor{gray!20}+191.5 \\ 
 & WB & \textbf{0.3} & 11.1 & 0.7 & +54.2 & 4.0 & 0.3 & -26.3 & 7.2 & 0.7 & +289.4 \\ 
\emph{\httppatchspring} & \cellcolor{gray!20}BB & \cellcolor{gray!20}0.0 & \cellcolor{gray!20}0.0 & \cellcolor{gray!20}\textbf{0.0} & \cellcolor{gray!20}-100.0 & \cellcolor{gray!20}0.0 & \cellcolor{gray!20}\textbf{0.0} & \cellcolor{gray!20}-100.0 & \cellcolor{gray!20}0.0 & \cellcolor{gray!20}\textbf{0.2} & \cellcolor{gray!20}-100.0 \\ 
 & WB & \textbf{2.8} & 38.8 & 0.5 & +1.8 & 38.8 & 0.5 & +1.8 & 0.0 & 0.5 & NaN \\ 
\emph{\languagetool} & \cellcolor{gray!20}BB & \cellcolor{gray!20}0.0 & \cellcolor{gray!20}0.0 & \cellcolor{gray!20}0.5 & \cellcolor{gray!20}NaN & \cellcolor{gray!20}0.0 & \cellcolor{gray!20}0.5 & \cellcolor{gray!20}NaN & \cellcolor{gray!20}0.0 & \cellcolor{gray!20}0.5 & \cellcolor{gray!20}NaN \\ 
 & WB & 0.0 & 0.0 & 0.4 & -100.0 & 0.0 & 0.5 & NaN & 0.0 & 0.4 & -100.0 \\ 
\emph{\market} & \cellcolor{gray!20}BB & \cellcolor{gray!20}\textbf{5.4} & \cellcolor{gray!20}15.1 & \cellcolor{gray!20}\textbf{0.0} & \cellcolor{gray!20}-29.6 & \cellcolor{gray!20}8.7 & \cellcolor{gray!20}0.3 & \cellcolor{gray!20}-15.0 & \cellcolor{gray!20}6.3 & \cellcolor{gray!20}\textbf{0.1} & \cellcolor{gray!20}-43.1 \\ 
 & WB & \textbf{12.0} & 16.2 & \textbf{0.2} & -23.2 & 9.8 & 0.3 & -15.4 & 6.3 & 0.2 & -32.8 \\ 
\emph{\microcks} & \cellcolor{gray!20}BB & \cellcolor{gray!20}\textbf{15.5} & \cellcolor{gray!20}16.0 & \cellcolor{gray!20}\textbf{0.8} & \cellcolor{gray!20}+16.9 & \cellcolor{gray!20}1.3 & \cellcolor{gray!20}\textbf{0.0} & \cellcolor{gray!20}-89.3 & \cellcolor{gray!20}14.7 & \cellcolor{gray!20}\textbf{1.0} & \cellcolor{gray!20}+1187.2 \\ 
 & WB & \textbf{74.8} & 1.5 & 0.5 & +0.3 & 1.5 & 0.5 & +0.3 & 0.0 & 0.5 & NaN \\ 
\emph{\ocvn} & \cellcolor{gray!20}BB & \cellcolor{gray!20}\textbf{6678.3} & \cellcolor{gray!20}23.1 & \cellcolor{gray!20}\textbf{0.0} & \cellcolor{gray!20}-19.3 & \cellcolor{gray!20}3.3 & \cellcolor{gray!20}\textbf{0.1} & \cellcolor{gray!20}-29.8 & \cellcolor{gray!20}19.8 & \cellcolor{gray!20}\textbf{0.0} & \cellcolor{gray!20}-17.2 \\ 
 & WB & \textbf{10357.6} & 46.8 & \textbf{0.0} & -15.3 & 1.6 & \textbf{0.0} & -63.1 & 45.2 & \textbf{0.0} & -11.3 \\ 
\emph{\ohsomeapi} & \cellcolor{gray!20}BB & \cellcolor{gray!20}0.0 & \cellcolor{gray!20}0.0 & \cellcolor{gray!20}0.5 & \cellcolor{gray!20}NaN & \cellcolor{gray!20}0.0 & \cellcolor{gray!20}0.5 & \cellcolor{gray!20}NaN & \cellcolor{gray!20}0.0 & \cellcolor{gray!20}0.5 & \cellcolor{gray!20}NaN \\ 
 & WB & 0.0 & 0.0 & 0.3 & -100.0 & 0.0 & 0.3 & -100.0 & 0.0 & 0.5 & NaN \\ 
\emph{\paypublicapi} & \cellcolor{gray!20}BB & \cellcolor{gray!20}0.0 & \cellcolor{gray!20}69.1 & \cellcolor{gray!20}0.6 & \cellcolor{gray!20}+3.7 & \cellcolor{gray!20}0.0 & \cellcolor{gray!20}0.5 & \cellcolor{gray!20}NaN & \cellcolor{gray!20}69.1 & \cellcolor{gray!20}0.6 & \cellcolor{gray!20}+3.7 \\ 
 & WB & 0.0 & 1.5 & 0.3 & -18.2 & 1.5 & 0.3 & -18.2 & 0.0 & 0.5 & NaN \\ 
\emph{\personcontroller} & \cellcolor{gray!20}BB & \cellcolor{gray!20}0.0 & \cellcolor{gray!20}31.7 & \cellcolor{gray!20}0.3 & \cellcolor{gray!20}-7.6 & \cellcolor{gray!20}31.7 & \cellcolor{gray!20}0.3 & \cellcolor{gray!20}-7.6 & \cellcolor{gray!20}0.0 & \cellcolor{gray!20}0.5 & \cellcolor{gray!20}NaN \\ 
 & WB & \textbf{4.3} & 23.8 & \textbf{0.0} & -20.3 & 23.8 & \textbf{0.0} & -20.3 & 0.0 & 0.5 & NaN \\ 
\emph{\proxyprint} & \cellcolor{gray!20}BB & \cellcolor{gray!20}\textbf{2472.0} & \cellcolor{gray!20}83.3 & \cellcolor{gray!20}\textbf{0.2} & \cellcolor{gray!20}-0.6 & \cellcolor{gray!20}75.0 & \cellcolor{gray!20}\textbf{0.0} & \cellcolor{gray!20}-10.5 & \cellcolor{gray!20}8.3 & \cellcolor{gray!20}\textbf{1.0} & \cellcolor{gray!20}+Inf \\ 
 & WB & \textbf{11.4} & 0.0 & \textbf{0.0} & -99.1 & 0.0 & \textbf{0.0} & -100.0 & 0.0 & 0.4 & -74.8 \\ 
\emph{\quartzmanager} & \cellcolor{gray!20}BB & \cellcolor{gray!20}0.0 & \cellcolor{gray!20}29.6 & \cellcolor{gray!20}0.5 & \cellcolor{gray!20}0.0 & \cellcolor{gray!20}7.4 & \cellcolor{gray!20}0.5 & \cellcolor{gray!20}0.0 & \cellcolor{gray!20}22.2 & \cellcolor{gray!20}0.5 & \cellcolor{gray!20}0.0 \\ 
 & WB & 0.0 & 16.1 & 0.6 & +22.9 & 16.1 & 0.6 & +22.9 & 0.0 & 0.5 & NaN \\ 
\emph{\reservationsapi} & \cellcolor{gray!20}BB & \cellcolor{gray!20}0.0 & \cellcolor{gray!20}24.8 & \cellcolor{gray!20}0.6 & \cellcolor{gray!20}+96.6 & \cellcolor{gray!20}12.6 & \cellcolor{gray!20}0.5 & \cellcolor{gray!20}0.0 & \cellcolor{gray!20}12.2 & \cellcolor{gray!20}0.6 & \cellcolor{gray!20}+Inf \\ 
 & WB & \textbf{41.6} & 45.2 & 0.5 & +19.2 & 10.0 & 0.4 & -11.0 & 35.2 & 0.5 & +31.8 \\ 
\emph{\restnews} & \cellcolor{gray!20}BB & \cellcolor{gray!20}\textbf{0.7} & \cellcolor{gray!20}26.3 & \cellcolor{gray!20}\textbf{0.1} & \cellcolor{gray!20}-34.9 & \cellcolor{gray!20}26.3 & \cellcolor{gray!20}0.3 & \cellcolor{gray!20}-19.4 & \cellcolor{gray!20}0.0 & \cellcolor{gray!20}\textbf{0.0} & \cellcolor{gray!20}-100.0 \\ 
 & WB & 0.0 & 3.7 & 0.5 & +0.4 & 3.7 & 0.5 & +0.4 & 0.0 & 0.5 & NaN \\ 
\emph{\restscs} & \cellcolor{gray!20}BB & \cellcolor{gray!20}0.0 & \cellcolor{gray!20}0.0 & \cellcolor{gray!20}0.5 & \cellcolor{gray!20}NaN & \cellcolor{gray!20}0.0 & \cellcolor{gray!20}0.5 & \cellcolor{gray!20}NaN & \cellcolor{gray!20}0.0 & \cellcolor{gray!20}0.5 & \cellcolor{gray!20}NaN \\ 
 & WB & 0.0 & 0.4 & 0.6 & +209.3 & 0.4 & 0.6 & +209.3 & 0.0 & 0.5 & NaN \\ 
\emph{\restcountries} & \cellcolor{gray!20}BB & \cellcolor{gray!20}0.0 & \cellcolor{gray!20}0.0 & \cellcolor{gray!20}0.5 & \cellcolor{gray!20}NaN & \cellcolor{gray!20}0.0 & \cellcolor{gray!20}0.5 & \cellcolor{gray!20}NaN & \cellcolor{gray!20}0.0 & \cellcolor{gray!20}0.5 & \cellcolor{gray!20}NaN \\ 
 & WB & 0.0 & 18.0 & 0.4 & -4.0 & 18.0 & 0.4 & -4.0 & 0.0 & 0.5 & NaN \\ 
\emph{\scoutapi} & \cellcolor{gray!20}BB & \cellcolor{gray!20}\textbf{75.4} & \cellcolor{gray!20}20.0 & \cellcolor{gray!20}\textbf{0.0} & \cellcolor{gray!20}-39.0 & \cellcolor{gray!20}14.9 & \cellcolor{gray!20}\textbf{0.1} & \cellcolor{gray!20}-50.0 & \cellcolor{gray!20}5.0 & \cellcolor{gray!20}0.6 & \cellcolor{gray!20}+77.4 \\ 
 & WB & \textbf{11.0} & 1.4 & 0.3 & -37.6 & 0.7 & 0.5 & -39.1 & 0.7 & 0.4 & -35.9 \\ 
\emph{\sessionservice} & \cellcolor{gray!20}BB & \cellcolor{gray!20}0.0 & \cellcolor{gray!20}0.0 & \cellcolor{gray!20}0.5 & \cellcolor{gray!20}NaN & \cellcolor{gray!20}0.0 & \cellcolor{gray!20}0.5 & \cellcolor{gray!20}NaN & \cellcolor{gray!20}0.0 & \cellcolor{gray!20}0.5 & \cellcolor{gray!20}NaN \\ 
 & WB & \textbf{1.0} & 0.0 & 0.4 & -100.0 & 0.0 & 0.4 & -100.0 & 0.0 & 0.5 & NaN \\ 
\emph{\springbatchrest} & \cellcolor{gray!20}BB & \cellcolor{gray!20}0.0 & \cellcolor{gray!20}0.0 & \cellcolor{gray!20}0.5 & \cellcolor{gray!20}NaN & \cellcolor{gray!20}0.0 & \cellcolor{gray!20}0.5 & \cellcolor{gray!20}NaN & \cellcolor{gray!20}0.0 & \cellcolor{gray!20}0.5 & \cellcolor{gray!20}NaN \\ 
 & WB & \textbf{29.3} & 22.0 & \textbf{0.1} & -29.3 & 21.0 & \textbf{0.1} & -32.5 & 1.0 & 0.6 & +Inf \\ 
\emph{\springecommerce} & \cellcolor{gray!20}BB & \cellcolor{gray!20}\textbf{2.0} & \cellcolor{gray!20}3.2 & \cellcolor{gray!20}\textbf{0.9} & \cellcolor{gray!20}+52.4 & \cellcolor{gray!20}3.2 & \cellcolor{gray!20}\textbf{1.0} & \cellcolor{gray!20}+97.4 & \cellcolor{gray!20}0.0 & \cellcolor{gray!20}0.3 & \cellcolor{gray!20}-100.0 \\ 
 & WB & \textbf{6.1} & 0.0 & \textbf{0.0} & -100.0 & 0.0 & \textbf{0.0} & -100.0 & 0.0 & 0.5 & NaN \\ 
\emph{\springrestexample} & \cellcolor{gray!20}BB & \cellcolor{gray!20}0.0 & \cellcolor{gray!20}1.6 & \cellcolor{gray!20}0.6 & \cellcolor{gray!20}+215.8 & \cellcolor{gray!20}1.6 & \cellcolor{gray!20}0.6 & \cellcolor{gray!20}+215.8 & \cellcolor{gray!20}0.0 & \cellcolor{gray!20}0.5 & \cellcolor{gray!20}NaN \\ 
 & WB & 0.0 & 1.3 & 0.3 & -61.9 & 1.3 & 0.3 & -61.9 & 0.0 & 0.5 & NaN \\ 
\emph{\swaggerpetstore} & \cellcolor{gray!20}BB & \cellcolor{gray!20}0.0 & \cellcolor{gray!20}43.0 & \cellcolor{gray!20}0.7 & \cellcolor{gray!20}+10.5 & \cellcolor{gray!20}34.3 & \cellcolor{gray!20}0.5 & \cellcolor{gray!20}-1.0 & \cellcolor{gray!20}8.7 & \cellcolor{gray!20}\textbf{0.8} & \cellcolor{gray!20}+102.1 \\ 
 & WB & \textbf{5.8} & 15.3 & 0.4 & -12.2 & 15.3 & 0.4 & -12.2 & 0.0 & 0.5 & NaN \\ 
\emph{\trackingsystem} & \cellcolor{gray!20}BB & \cellcolor{gray!20}\textbf{62.8} & \cellcolor{gray!20}59.1 & \cellcolor{gray!20}\textbf{0.0} & \cellcolor{gray!20}-20.6 & \cellcolor{gray!20}40.8 & \cellcolor{gray!20}\textbf{0.0} & \cellcolor{gray!20}-42.0 & \cellcolor{gray!20}18.3 & \cellcolor{gray!20}\textbf{1.0} & \cellcolor{gray!20}+346.6 \\ 
 & WB & \textbf{43.0} & 45.3 & 0.4 & -3.2 & 42.8 & 0.4 & -1.1 & 2.5 & \textbf{0.2} & -28.7 \\ 
\emph{\usermanagement} & \cellcolor{gray!20}BB & \cellcolor{gray!20}\textbf{6.5} & \cellcolor{gray!20}27.4 & \cellcolor{gray!20}\textbf{0.0} & \cellcolor{gray!20}-29.2 & \cellcolor{gray!20}25.2 & \cellcolor{gray!20}\textbf{0.0} & \cellcolor{gray!20}-34.3 & \cellcolor{gray!20}2.3 & \cellcolor{gray!20}\textbf{0.9} & \cellcolor{gray!20}+425.5 \\ 
 & WB & \textbf{8.5} & 2.8 & \textbf{0.1} & -68.0 & 2.8 & \textbf{0.1} & -68.0 & 0.0 & 0.5 & NaN \\ 
\emph{\webgoat} & \cellcolor{gray!20}BB & \cellcolor{gray!20}\textbf{19.9} & \cellcolor{gray!20}89.6 & \cellcolor{gray!20}\textbf{1.0} & \cellcolor{gray!20}+144.8 & \cellcolor{gray!20}9.3 & \cellcolor{gray!20}\textbf{0.0} & \cellcolor{gray!20}-74.4 & \cellcolor{gray!20}80.3 & \cellcolor{gray!20}\textbf{1.0} & \cellcolor{gray!20}+25953.0 \\ 
 & WB & 0.0 & 0.4 & 0.5 & +0.0 & 0.4 & 0.5 & +0.0 & 0.0 & 0.5 & NaN \\ 
\emph{\youtubemock} & \cellcolor{gray!20}BB & \cellcolor{gray!20}\textbf{0.8} & \cellcolor{gray!20}1.7 & \cellcolor{gray!20}0.5 & \cellcolor{gray!20}+84.6 & \cellcolor{gray!20}1.7 & \cellcolor{gray!20}0.5 & \cellcolor{gray!20}+84.6 & \cellcolor{gray!20}0.0 & \cellcolor{gray!20}0.5 & \cellcolor{gray!20}NaN \\ 
 & WB & 0.0 & 0.0 & 0.5 & NaN & 0.0 & 0.5 & NaN & 0.0 & 0.5 & NaN \\ 
\midrule 
\multicolumn{2}{l}{\textbf{Summary-BB}} & 9343.2 & 30.6 & -- & -- & 18.2 & -- & -- & 12.4 & -- & -- \\ 
\multicolumn{2}{l}{\textit{Handled APIs}} & 13 & \multicolumn{9}{l}{} \\ 
\multicolumn{2}{l}{\textit{Reduced APIs}} & -- & -- & 12 (10) & -- & -- & 15 (9) & -- & -- & 7 (4) & -- \\ 
\multicolumn{2}{l}{\textit{Reduced to 0}} & -- & -- & 0 & -- & -- & 0 & -- & -- & 3 & -- \\ 
\multicolumn{2}{l}{\textbf{Summary-WB}} & 10631.4 & 16.3 & -- & -- & 10.8 & -- & -- & 5.5 & -- & -- \\ 
\multicolumn{2}{l}{\textit{Handled APIs}} & 18 & \multicolumn{9}{l}{} \\ 
\multicolumn{2}{l}{\textit{Reduced APIs}} & -- & -- & 18 (8) & -- & -- & 19 (7) & -- & -- & 7 (2) & -- \\ 
\multicolumn{2}{l}{\textit{Reduced to 0}} & -- & -- & 2 & -- & -- & 3 & -- & -- & 9 & -- \\ 
\bottomrule 
\end{tabular} 

	}
	\vspace{-0.5cm}
\end{table}

Table~\ref{tab:flakiness_handled} shows that \method is effective in both BB and WB settings. 
For example, on \ocvn, our approach mitigates on average 6678.3 flaky assertions with BB and 10357.6 with WB on average over 10 generations. 
Across all 31 flaky APIs, \method resolves, on average, 10631.4 flaky assertions with WB and 9343.2 with BB. 
It also handles flaky assertions in 18 APIs with WB, compared with 13 APIs with BB. 


In Table~\ref{tab:flakiness_handled}, we also report the overall failure rate ($FR$\%), the rate of consistent failures ($FR_c = \#F_c / \#F$\%), the rate of unstable failures ($FR_u = \#F_u / \#F$\%), together with comparisons against tests generated without handling using $\hat{A}_{12}$ and relative improvement.
In terms of \textit{FR}\%, with BB, the remaining flaky rates are reduced to $FR_c=18.2\%$ and $FR_u=12.4\%$, while with WB they further decrease to $10.8\%$ and $5.5\%$, respectively.
Moreover, with the WB setting, our approach reduced $FR$ in 18 of the 31 APIs, with 8 of these reductions being statistically significant (i.e., $\hat{A}_{12} < 0.5$ and $p < 0.05$), and completely resolved 2 APIs (i.e., \proxyprint and \springecommerce). 
With the BB setting, our approach reduced $FR$ in 12 APIs, 10 of which showed statistically significant reductions.
For consistent failures, the approach reduces $FR_c$ in 15 APIs in BB and 19 APIs in WB.
Among these reductions, 9 in BB and 7 in WB are statistically significant. 
Moreover, the remaining consistent-failure rate is reduced to zero in 3 APIs with WB, whereas no API is reduced to zero in BB. 
For unstable failures, the approach reduces $FR_u$ in 7 APIs in both BB and WB. 
However, WB eliminates unstable failures completely in 9 APIs, compared with only 3 in BB, although only 2 WB reductions and 4 BB reductions are statistically significant. 

There are also cases where the failure rate increases, as observed in \featuresservice, \gestaohospital, \microcks, \springecommerce, and \webgoat. 
This suggests that some forms of flakiness cannot be effectively resolved without access to source code and internal state.

\begin{results}[Findings of RQ3]
\method effectively reduces flaky tests in both BB and WB settings, with better overall performance in WB.
It reduces both consistent and unstable failures, and can completely resolve flakiness for some APIs.
However, its effectiveness varies across APIs, and some residual flakiness remains difficult to address without access to source code and internal state.
\end{results}

\section{Threats To Validity}
\label{sec:threats}

\textit{Construct validity.} Our taxonomy of flakiness sources in REST API testing is derived from a manual review of near 3000 of failing tests. This process may introduce human errors and subjective bias. To mitigate this threat, each classification was independently assessed by two authors, with each disagreement resolved through discussions.

\textit{Internal validity.} Due to the inherent randomness of fuzzing, experimental results may vary across runs. To mitigate this threat, we repeated each experiment 10 times and analyzed observations across repetitions.

\textit{External validity.} Our empirical evaluation is based on a dataset of 36 open-source APIs. While relatively large, this sample is by no means representative of all APIs, particularly those developed and deployed in industrial settings. Therefore, we acknowledge that the generalizability of our findings may be limited.

\textit{Conclusion validity.} Our study focuses on test cases generated using a single fuzzer: \evo. As different fuzzing or test generation techniques may exhibit different characteristics, our findings may not directly generalize to other approaches. Nevertheless, our observations provide empirical evidence that can serve as a baseline for comparison in future studies. To support this, we make all collected data publicly available, including the manually labeled failing test cases, enabling replication and further investigation by the research community.

\section{Conclusions}
\label{sec:conclusions}
In this paper, we presented, to the best of our knowledge, the first systematic study of flakiness in test cases generated by fuzzing techniques for REST APIs. 
Our empirical study on 36 APIs, based on test cases generated by \evo under both black-box and white-box settings, led to the identification of a taxonomy comprising nine distinct sources of flakiness. Building on this taxonomy, we designed and evaluated \method to detect and mitigate flakiness in tests generated by white-box and black-box fuzzers for REST APIs. 
As future work, we plan to extend our analysis to test cases generated by other techniques and to develop more advanced detection and mitigation strategies to reduce the impact of flakiness while preserving test effectiveness.

\section*{Acknowledgments}
This work is supported by the National Science Foundation of China (grant agreement No. 62502022).
Andrea Arcuri is funded by the European Research Council (ERC) under the European Union’s Horizon 2020 research and innovation programme (EAST project, grant agreement No. 864972).

\section*{Data Availability Statement}
All our analysis results, and code extension to \evo to handle flakiness, are available at: \url{https://anonymous.4open.science/r/FlakyCatch-8312}.

%
%
%
%
%

\bibliographystyle{ACM-Reference-Format}

\end{document}